\documentclass[12pt]{iopart}
\usepackage{iopams}
\usepackage{epsf}
\newif\ifpdf
  \ifx\pdfoutput\undefined
     \pdffalse
\else
  \pdfoutput=1
  \pdftrue
\fi

\ifpdf
\usepackage{times}
\usepackage[pdftex,
        pdfstartview=FitV,
        colorlinks=true,
        urlcolor=rltblue,       
        filecolor=rltgreen,     
        linkcolor=rltred,       
        citecolor=blue,     
        pagebackref,
        bookmarksopen=true]{hyperref}
\usepackage{color}
\definecolor{rltred}{rgb}{0.75,0,0}
\definecolor{rltgreen}{rgb}{0,0.5,0}
\definecolor{rltblue}{rgb}{0,0,0.75}
\usepackage[pdftex]{graphicx}
\else
\usepackage[dvips]{graphicx}
\usepackage{color}
\fi
\begin{document}

\title[Modelisation of transition and noble metal vicinal surfaces]
{Modelisation of transition and noble metal vicinal surfaces: energetics,
vibrations and stability}

\author{C. Barreteau$^*$, F. Raouafi$^*$,  M.C. Desjonqu\`eres$^*$
and  D. Spanjaard$^{\dag}$}
\address{
$^*$DSM/DRECAM/SPCSI, CEA Saclay
\newline F-91 191 Gif sur Yvette, France \newline
$^{\dag}$Laboratoire de Physique des Solides, Universit\'e
Paris Sud, F-91 405 Orsay, France \newline
}
\date {\today}

\maketitle

\begin{abstract}
The energetics of transition and noble metal (Rh, Pd, Cu) vicinal
surfaces, i.e., surface energy, step energy, kink energy and
electronic interactions between steps, is studied at 0K from
electronic structure calculations in the tight-binding
approximation using a {\it s, p} and {\it d} valence orbital basis
set. Then, the surface phonon spectra of copper are investigated
in the harmonic approximation with the help of a semi-empirical
inter-atomic potential. This allows to derive the contribution of
phonons at finite temperatures to the step free energy and to the
interactions between steps. The last part is devoted to the
stability of vicinal surfaces relative to faceting with special
attention to the domain of orientations (100)-(111).
Semi-empirical potentials are shown to be not realistic enough to
give a reliable answer to this problem. The results derived from
electronic structure calculations predict a variety of behaviors
and, in particular, a possible faceting into two other vicinal
orientations. Finally, temperature effects are discussed.
Comparisons are made with other theoretical works and available
experiments.

\end{abstract}

\thispagestyle{empty}
\section{Introduction}

Studies of vicinal surfaces of metals have given rise to numerous
experimental and theoretical works. Indeed the role of steps and
kinks is fundamental for understanding the morphology of crystal
surfaces and, in particular, its evolution with time and
temperature as well as the equilibrium surface structure. In
addition, vicinal surfaces may provide appropriate substrates for
growing nanostructures, for instance nanowires, with magnetic and
transport properties of high technological interest.

In the last twenty years the direct investigation of the local
surface structure has become possible by the use of scanning probe
microscopies such as the Scanning Tunneling Microscope (STM).
Information on the energetics of surface defects can henceforth be
derived from a statistical study of STM images and their evolution
with temperature. For instance, the study of the equilibrium shape
of large adislands grown in homoepitaxy on monocrystalline
surfaces has been used to determine the anisotropy of step
energies and, more recently, the absolute values of step and kink
energies \cite{Giesen01}. Furthermore, the interaction between
steps can be deduced from the study of terrace width distributions
\cite{Pai94}. Kink energies can also be obtained from the
observation of the spatial equilibrium fluctuations of step edges
\cite{Villain85,Barbier96}. In addition experimental
investigations of localized vibrational modes at vicinal surfaces
have been carried out in the last decade by Inelastic Helium Atom
Scattering (IHAS)\cite{Witte95} or Electron Energy Loss
Spectroscopy (EELS)\cite{Kara00}.

All these experimental results have motivated a lot of theoretical
works. The study of the energetics of vicinal surfaces at 0K
starts from the determination of the surface energies as a
function of the surface orientation. It has been investigated
either by using semi-empirical potentials including an N-body
contribution such as Effective Medium Theory
(EMT)\cite{Jacobsen87,Stoltze94}, Embedded Atom Method
(EAM)\cite{Daw84}, Second Moment Potential
(SMA)\cite{Ducastelle70,Sutton84,Finnis84}, or starting from the
determination of the electronic structure using the Density
Functional Theory (DFT) or the Tight Binding approximation (TB).
However, due to the very low symmetry of these systems, first
principle calculations are scarce and limited to a very small
number of geometries and metals: Al \cite{Nelson92,Stumpf96}, Cu
\cite{Feibelman99,Spisak01}, Pt \cite{Feibelman00}. On the
contrary, in the case of transition metals, TB methods are able to
describe correctly the quantum mechanical effects without a lot of
computational efforts. Using this method we have been able
\cite{Raouafi102} to perform a systematic study of various vicinal
surfaces of Rh, Pd and Cu as a function of the misorientation
angle from which we have deduced step and kink energies as well as
step-step interactions. The results of this work are reviewed in
Sect.3, after a brief presentation of the geometry of vicinal
surfaces (Sect.2).

Similarly ab-initio methods have also been used to obtain
localized vibration modes at vicinal surfaces but only the modes
at high symmetry points of the Brillouin zone have been
investigated \cite{Chen91,Wei98}. Central pair potentials in the
harmonic approximation yield a reasonable description of vibration
modes but not at a quantitative level. A good accuracy can be
achieved with N-body semi-empirical potentials such as EAM
\cite{Kara96,Sklyadneva98}. More recently we have set up a new
potential for Cu \cite{Barreteau02} with which we have been able
to reproduce accurately the phonon dispersion curves measured by
IHAS \cite{Witte95} and EELS \cite{Kara00} on
flat as well as on vicinal surfaces. In Sect.4 this
potential is presented and the results concerning phonon
dispersion curves and the vibrational contribution to the step
free energy are summarized.

Finally, it is not obvious that all vicinal surfaces should be in
thermodynamic equilibrium. Actually vicinal surfaces may decrease
their surface free energy by rearranging the atoms in order to
exhibit a hill and valley (or a factory roof) structure. This
phenomenon, called faceting, is indeed observed in experiments and
its occurrence can be predicted from the knowledge of the spatial
anisotropy of the surface energy \cite{Herring51}. Using the
results of Sects.3 and 4. we have been able to reexamine
\cite{Desjonqueres02,Raouafi202} in a realistic way this old
problem. In Sect.5 we report our main results and discuss the
implications of doing electronic structure calculations rather
than using empirical potentials.

\section{The geometry of vicinal surfaces}

\begin{figure}[!fht]
\begin{center}
  \ifpdf
  \includegraphics*[width=15cm]{figure1.pdf}
  \else
  \includegraphics*[width=15cm]{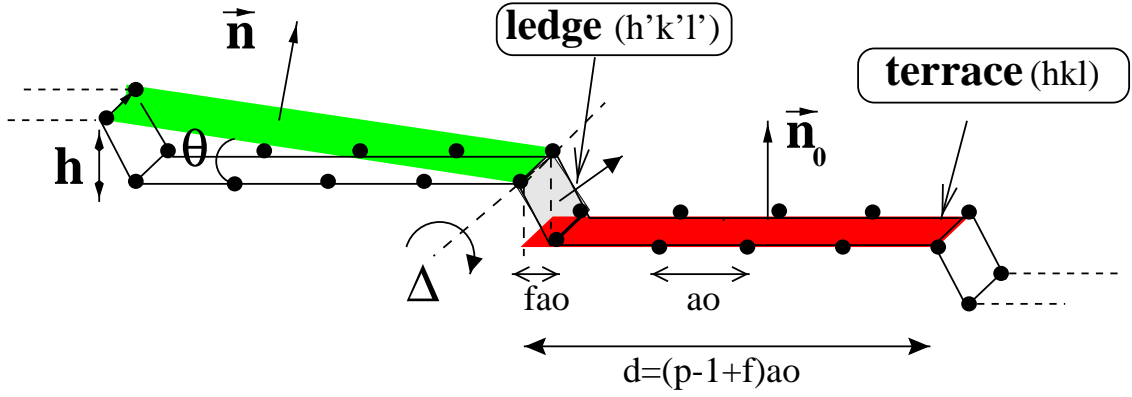}
  \fi
\caption{Step geometry of a $p(hkl)\times (h'k'l')$ vicinal surface}
\label{fig:cut}	  
\end{center}
\end{figure}

A vicinal surface is obtained by cutting a crystal along a plane
making an angle $\theta$ with respect to a low index plane normal
to the direction ${\bf n_0}$ (Fig.\ref{fig:cut}). For selected values of
$\theta$, such a surface can be viewed as a periodic succession of
terraces normal to ${\bf n_0}$, with equal widths, separated by steps
of monoatomic height. The width of the terraces is determined by
the number $p$ of atomic rows (including the inner edge) parallel
to the step edges. A vicinal surface corresponds to an atomic
plane with high Miller indices. It can also be denoted using the
Lang et al. \cite{Lang72} notation $p(hkl)\times(h'k'l')$, $(hkl)$ and
$(h'k'l')$ being respectively the Miller indices of planes
parallel to the terraces and to the ledges. Note also that when
projecting the unit cell of the vicinal surface on the terrace
plane, a geometrical factor $f$ occurs when the ledges and the
terraces are not orthogonal.

In the following we consider FCC crystals and four step geometries
with (111) and (100) terraces. The geometrical features of the
four types of vicinal surfaces, denoted using Lang et al.
notations and Miller indices, are given in Table \ref{tab:miller_indices}.
 One can note
that for a given step geometry there often exists two types of
unit cell (primitive rectangular and centered rectangular)
depending on the width of the terrace (p even or odd). In the
first three considered geometries, the atoms along the step edges
are first nearest neighbors while for the $p(100)\times(010)$
surface the atoms are second nearest neighbors and, consequently,
the corresponding step edge has a zigzag shape that can be seen as
a succession of kinks.

\begin{table}
\begin{center}
\begin{tabular}{c|c|c|c|c}
Lang et al. notations         & Miller indices   &  f    &  Edge geometry & 2D unit cell \\
\hline
$p(111)\times(100)$ step A & $(p+1,p-1,p-1)$  & $2/3$ &       nn       &  p odd: PR   \\
                           &                  &       &                &  p even: CR  \\
$p(111)\times(\bar{1}11)$ step B & $(p-2,p,p)$      & $1/3$ &       nn       &  p odd: CR   \\
                           &                  &       &                &  p even:PR   \\
$p(100)\times(111)$        & $(1,1,2p-1)$     & $1/2$ &       nn       &      CR      \\
$p(100)\times(010)$        & $(0,1,p-1)$      & $0$   &      nnn       &  p odd: CR   \\
                           &                  &       &                &  p even: PR
\end{tabular}
\caption{ Geometrical features of the four types of vicinal
surfaces. The geometry of the step edges is indicated by the
distance between two consecutive atoms: nearest neighbors (nn),
next nearest neighbors (nnn). The nature of the 2D unit cell is
rectangular, either primitive (PR) or centered (CR). Finally the
usual notations, step A and step B, for the vicinals of (111) are
indicated }
\label{tab:miller_indices}
\end{center}
\end{table}

\section{Electronic structure and energetics of vicinal surfaces in a tight-binding model.}

\subsection{The spd tight-binding model.}

The study of the electronic structure of vicinal surfaces has been
carried out using a slab model. The orientation of the normal to
the vicinal surface is first chosen. A succession of $N_{slab}$
atomic layers is built, $N_{slab}$ being large enough so that the
interaction between the two free surfaces of the slab is
negligible. The wider the terraces (i.e., the area of the vicinal
surface unit cell), the smaller the inter-layer spacing and the
larger the number of layers $N_{slab}$. The system has thus a
two-dimensional periodicity with $N_{slab}$ atoms per unit cell.
Ab-initio calculations are in principle feasible. However they
are, at least up to now, limited to small terrace widths since
they need a large amount of computer time. On the contrary the TB
scheme is very attractive since it is much less costly in computer
time and still describes systems within the framework of quantum
mechanics. Up to the last decade, the TB basis set was limited to
the valence {\it d} $(xy,yz,zx,x^2-y^2,3z^2-r^2)$ orbitals and
this scheme was quite successful to explain the general trends in
the variation of a large number of physical properties along the
transition series. However for FCC elements at the end of these
series, the values of energetic quantities are then
underestimated, and even cancel for a full {\it d} band, due to
the neglect of the contribution of the outer $s$ and $p(x,y,z)$
orbitals. Recently, it was found that it is possible to determine
a transferable parametrized TB hamiltonian in a {\it spd} basis
set giving not only a quite good description of the band structure
up to a few eV above the {\it d} band, but also total energies
with a good accuracy. This was initially proposed by Mehl and
Papaconstantopoulos who assumed a non orthogonal basis set
\cite{Mehl96}. Then it was shown \cite{Barreteau98} that for
elements with a not completely filled {\it d} band, it was
possible to reduce considerably the number of parameters by
assuming an orthogonal basis set.

These models have been described in details in Refs
\cite{Mehl96,Barreteau98} thus their main features are only
briefly recalled in the following. The interatomic matrix elements
of the hamiltonian $H^{\lambda\mu}_{ij}$ (i,j: atomic sites,
$\lambda,\mu$: atomic orbitals) in the two-center approximation are
determined from the ten Slater-Koster (SK) \cite{Slater54} hopping integrals
$ss\sigma, sp\sigma, sd\sigma, pp\sigma, pp\pi, pd\sigma, pd\pi,
dd\sigma, dd\pi, dd\delta$. The laws of variation with distance of
the SK hopping integrals are a simple exponential decay in
Ref.\cite{Barreteau98} and are slightly more involved with a larger
number of parameters in Ref.\cite{Mehl96}. In the non-orthogonal
scheme the overlap integrals $S^{\lambda\mu}_{ij}$  introduce also ten 
SK-like overlap parameters and their variations with distance
follow the same kind of laws as the interatomic matrix elements of
$H$. Following Ref.\cite{Mehl96}, in both schemes the intra-atomic
matrix elements $H^{\lambda\lambda}_{ii}$, i.e., the atomic levels
$\varepsilon_s,\varepsilon_p,\varepsilon_d$, are defined in such a
way that the total energy is obtained by summing up the occupied
energy levels. This means that all the other terms contained in the
energy functional of the DFT have been taken into account by a
rigid shift of the bulk band structure. As a consequence the
atomic levels should depend on the atomic environment and are
written in the form:

\begin{equation}
\varepsilon_{i\lambda}^0=a_{\lambda}+b_{\lambda}\rho_i^{2/3}+c_{\lambda}\rho_i^{4/3}
+d_{\lambda}\rho_i^{2}
\end{equation}

\noindent with

\begin{equation}
\rho_i=\sum_{j \neq i} exp(-p_{\rho}(R_{ij}/R_0 - 1))
\label{rho_i}
\end{equation}

\noindent where $R_{ij}$ is the distance between atoms i and j and
$R_0$ is a reference distance, usually the bulk equilibrium
interatomic spacing.

The parameters of the model are determined by a non-linear least mean square
fit on ab-initio band structure and total energy
for a few crystallographic bulk structures (usually FCC and BCC) at
several interatomic distances. Their values for Palladium and
Rhodium have been given in Ref.\cite{Barreteau98,Barreteau00}
and for Copper in Ref.\cite{Mehl96,Mehlweb}.

It should be noted that these parameters are obtained from systems
in which all atoms are neutral since they are
geometrically equivalent. When this is not the case we have added
a shift $\delta V_i$ to the on-site terms in order
to ensure local charge neutrality which should be almost strictly obeyed
in metals. Note that in the non-orthogonal case this induces also a modification
$\delta V_{ij}^{\lambda\mu}$ of the interatomic elements of $H$ \cite{Raouafi102}.
These potentials
arise from electron-electron interactions, thus one should
subtract the corresponding double counting terms
from the sum of occupied levels in the
expression of the total energy which is then written in both schemes as:

\begin{equation}
E_{tot}= \sum_{n occ} \epsilon_n -N_{val} \sum_i \delta V_i
\label{Edc}
\end{equation}

\noindent where $N_{val}$ is the total number of valence $spd$
electrons per atom of the metal.

\subsection{The electronic structure}

In order to apply the two-dimensional Bloch theorem, a basis set
of 2D Bloch waves localized in each layer {\it l} is defined as follows:

\begin{equation}
B_{l\lambda}({\bf r},{\bf k}_{//})= N_S^{-1/2} \sum_{i\in l}
exp(i{\bf k}_{//}.{\bf R}_{i//})\vert i\lambda>
\end{equation}

\noindent where $N_S$ is the number of atoms in each layer and
${\bf R}_{i//}$ the translation vectors of the two-dimensional (2D) lattice. Using
this basis set the solution of the Schr\"odinger equation can be
written:

\begin{equation}
\Psi_{{\bf k}_{//},n}({\bf r})= \sum_{l\lambda}c^n_{l\lambda}({\bf k}_{//})
B_{l\lambda}({\bf r},{\bf k}_{//})
\end{equation}

\noindent and the hamiltonian and overlap matrices are reduced to
$(9N_{slab}\times9N_{slab})$ matrices $H^{\lambda\mu}_{ll'}$ (${\bf k}_{//}$)
and $S^{\lambda\mu}_{ll'}({\bf k}_{//})$. Then the equation:

\begin{equation}
\sum_{l'\mu} [H^{\lambda\mu}_{ll'}({\bf k}_{//})-\epsilon_n({\bf
k}_{//}) S^{\lambda\mu}_{ll'}({\bf k}_{//})] c^n_{l'\mu}({\bf
k}_{//})=0
\end{equation}

\noindent is solved. Note that when the basis is orthonormal this
equation reduces to a classical eigenvalue(vector) problem. In
order to determine the projected band structure $\epsilon_n({\bf
k}_{//})$, ${\bf k}_{//}$ is varied along symmetry lines of the
surface Brillouin zone (SBZ). Other interesting quantities can be
calculated such as the local density of states (LDOS) at layer $l$
(per surface atom).

\begin{equation}
n_l(E)=\sum_{\lambda, n \atop l',\mu}
\frac{A}{(2\pi)^2}\int_{SBZ}c^{n*}_{l\lambda}({\bf k}_{//}) 
S^{\lambda\mu}_{ll'}({\bf k}_{//}) c^n_{l'\mu}({\bf k}_{//})
 \delta(E-\epsilon_n)d^2{\bf
k}_{//}
\end{equation}

\noindent where A is the area of the surface unit cell, or the
spectral local DOS (per surface atom)

\begin{equation}
n_l(E,{\bf k}_{//})=\sum_{\lambda, n \atop l',\mu} c^{n*}_{l \lambda}({\bf k}_{//}) 
S^{\lambda \mu}_{ll'}({\bf k}_{//}) c^n_{l'\mu}({\bf k}_{//}) \delta(E-\epsilon_n)
\end{equation}

\noindent corresponding to a given value of ${\bf k}_{//}$.

Some typical examples of surface projected densities of states,
$n_l(E)$ and $n_l(E,{\bf k}_{//})$ have been given in
Ref.\cite{Raouafi102} for several vicinal surfaces of Rhodium. The
most striking feature is the disappearance of almost all gaps.
Indeed, as the width of the terraces increases, the area of the SBZ
decreases and the height of the surface
adapted Bulk Brillouin zone (BBZ), i.e., the
sampled domain of bulk states (corresponding to all possible
values of $k_z$) increases accordingly and corresponds to lines
with no symmetry in the BBZ. This explains the absence of gaps and
of true surface states. However a number of resonances can be
identified. When the terrace width tends to infinity the spectral
DOS becomes vanishingly small in the energy domain corresponding
to gaps in the projected band structure of the flat surface with
the same orientation as the terraces.

\subsection{Surface and step energies}

The calculation of the total energy of the slab, from which
surface and step energies are deduced, involves a summation over
the SBZ which is carried out by using special ${\bf k}_{//}$
points belonging to the irreducible part of the SBZ
\cite{Cunningham74}, each energy level being broadened by the
derivative of a Fermi function of width $w_f$. The surface energy
per surface atom of the vicinal surface is obtained from the
following equation:

\begin{equation}
E_S({\bf n})=\frac{E_{slab}({\bf n})-N_{slab}E_{bulk}}{2}
\end{equation}

\noindent where $E_{slab}$ is the total energy of the slab (with
$N_{slab}$ layers) per surface unit cell and $E_{bulk}$ is the
energy of a bulk atom. The corresponding surface energy per unit
area is thus $\gamma({\bf n})=E_S({\bf n})/A({\bf n})$ where
$A({\bf n})$ is the area of the surface unit cell.

The step energy per unit step length $\beta(\theta)$ of a vicinal surface is
usually defined by the formula:

\begin{equation} \label{eq:gamma}
\gamma ({\bf n}) = \gamma ({\bf n_0}) cos(\theta) + \beta (\theta) sin(\theta)/h
\end{equation}

\noindent where $h$ is the inter-planar distance along the
direction ${\bf n_0}$ normal to the terraces. Note that due to the
presence of the array of steps with a period depending on
$\theta$, $\beta (\theta)$ is expected to vary with $\theta$ as a
result of step-step interactions. The value of the step energy for
an isolated step is then obtained in the limit $\theta \rightarrow
0$. It is easy to show \cite{Vitos99} that equation
(\ref{eq:gamma}) can be transformed into a more convenient form:

\begin{equation} \label{eq:estep}
E_{step}({\bf n_0},p) = E_S({\bf n_0},p) - (p-1+f)E_S({\bf
n_0},\infty)
\end{equation}

\noindent where $E_{step}({\bf n_0},p)$ is now the step energy per step atom
of the vicinal surface in which the terraces of orientation ${\bf
n_0}$ have $p$ atomic rows parallel to the step edge (including
the inner edge) and $E_S({\bf n_0},p)$ ($E_S({\bf n_0},\infty)$) is the surface energy
per surface atom of the vicinal (flat) surface. Finally, $f$ is a
geometrical factor depending on the vicinal surface which has been
defined in Sect.2 (Fig.1).

The calculation of step energies and especially of their variation
with $p$ is rather tricky since the step energies are of the order
of a few $10^{-1}$ eV and the magnitude of their variation with
$p$ is, at most, $\simeq 2.10^{-2}$eV. Thus the surface energies
involved in (\ref{eq:estep}) must be calculated with an accuracy
of $10^{-3}$eV. The parametrized TB hamiltonian being given, the
accuracy of the calculation is mainly governed by the thickness of
the slab, the number of ${\bf k}_{//}$ points in the irreducible
part of the SBZ and the Fermi level broadening $w_f$. Note that we
have extrapolated the total energy at zero broadening using the
usual approximation of Ref.\cite{Weinert92}:

\begin{equation}
E_{tot}(T=0) \approx E_{tot}(T) - \frac{1}{2} T S_e  + O(T^2)
\end{equation}

\noindent where $S_e$ is the electronic entropy and $T$ is the
electronic temperature corresponding to the Fermi broadening
$w_f$. We have found that the required accuracy is achieved when
using a number of vicinal planes in the slab $N_{slab}=pn_{slab}$
with $n_{slab} \simeq 10$, 64 special ${\bf k}_{//}$ points and a
Fermi level broadening of 0.2eV. Furthermore the
iteration process ensuring the self-consistent charge neutrality
condition has been stopped when the difference of charge between
two consecutive iterations is $<0.01e^{-}$ per atom and the
difference in total energy smaller than $10^{-4}$eV.

\begin{table}
\begin{center}
\begin{tabular}{c|c|c|c|c}
   &  \multicolumn{4}{c}{Rh (a=3.81{\AA})} \\
\hline
   & \multicolumn{2}{c|}{$p(100)\times(111)$}   &  \multicolumn{2}{c}{$p(111)\times(100)$} \\
\hline
p  &  $\gamma$ & $E_S$ & $\gamma$ & $E_S$ \\
\hline
2  &  3.281    &  2.465   &  3.281    &  2.465     \\
3  &  3.259    &  3.836   &  3.200    &  3.551     \\
4  &  3.225    &  5.217   &  3.120    &  4.634     \\
5  &  3.197    &  6.597   &  3.066    &  5.726     \\
6  &  3.175    &  7.976   &  3.026    &  6.819     \\
7  &  3.158    &  9.355   &  2.995    &  7.911     \\
$\infty$ & 3.044 &  1.379   &  2.781  &  1.091     \\
\hline \hline
   &  \multicolumn{4}{c}{Pd (a=3.89{\AA})} \\
\hline
   & \multicolumn{2}{c|}{$p(100)\times(111)$}   &  \multicolumn{2}{c}{$p(111)\times(100)$} \\
\hline
p  &  $\gamma$ & $E_S$ & $\gamma$ & $E_S$ \\
\hline
2  &  1.957    &  1.533   &  1.957    &  1.533     \\
3  &  1.922    &  2.358   &  1.897    &  2.194     \\
4  &  1.890    &  3.188   &  1.847    &  2.860     \\
5  &  1.867    &  4.016   &  1.811    &  3.526     \\
6  &  1.850    &  4.845   &  1.784    &  4.191     \\
7  &  1.837    &  5.673   &  1.764    &  4.857     \\
$\infty$  &  1.754   &  0.828  &  1.625 &  0.665   \\
\hline \hline
   &  \multicolumn{4}{c}{Cu (a=3.52{\AA})} \\
\hline
   & \multicolumn{2}{c|}{$p(100)\times(111)$}   &  \multicolumn{2}{c}{$p(111)\times(100)$} \\
\hline
p  &  $\gamma$ & $E_S$ & $\gamma$ & $E_S$ \\
\hline
2  &  2.049    &  1.314   &  2.049    &  1.314     \\
3  &  2.051    &  2.060   &  2.000    &  1.895     \\
4  &  2.034    &  2.809   &  1.953    &  2.476     \\
5  &  2.019    &  3.557   &  1.917    &  3.057     \\
6  &  2.008    &  4.306   &  1.892    &  3.639     \\
7  &  1.999    &  5.054   &  1.872    &  4.220     \\
$\infty$  &  1.935  &  0.748  &  1.734  &  0.581   \\
\end{tabular}
\caption{The surface energies $\gamma(J/m^2)$ and $E_S(eV/atom)$
of two families of vicinal surfaces. The surface area of the unit
cell of a $p(100)\times(111)$ surface is $S=\sqrt{(2p-1)^2+2} \, \, a^2/4$ (a: lattice parameter) 
and the angle $\theta$ is given by $\tan \theta=\sqrt{ 2}/(2p-1)$. The corresponding quantities for the
$p(111)\times(100)$ surface are: $S=\sqrt{(p+1)^2+2(p-1)^2}\, \,a^2/4$ and $\tan \theta=2\sqrt{2}/(3p-1)$.}
\label{tab:energ_surf}
\end{center}
\end{table}

\begin{figure}[!fht]
\begin{center}
  \ifpdf
  \includegraphics*[width=15cm]{figure2.pdf}
  \else
  \includegraphics*[width=15cm]{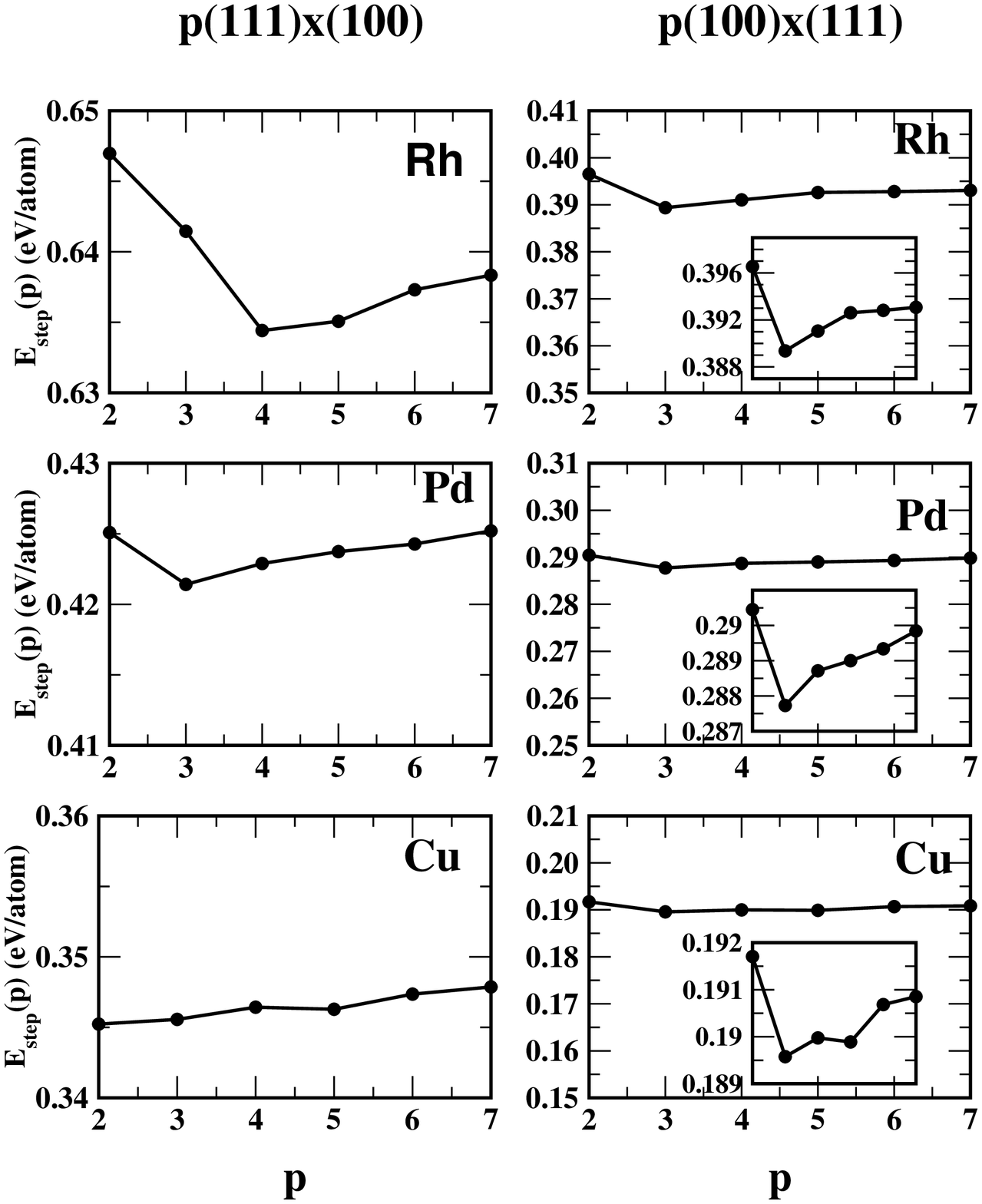}
  \fi
\caption{The variation of the step energy (per step atom) as a
function of the terrace width for the
$p(100)\times(111)$ (for which the energy scale has been enlarged
in the insets to put forward clearly the sign of interactions) and
$p(111)\times(100)$ vicinal surfaces of Rh, Pd and Cu.}
\label{fig:estepp}	  
\end{center}
\end{figure}

In Table \ref{tab:energ_surf} the surface energies per unit surface 
area $\gamma ({\bf n})$ and $E_S({\bf n_0},p)$
per surface atom are given as a function of $p$ for the
$p(100)\times(111)$ and $p(111)\times(100)$ surfaces. The step
energies per step atom are deduced from (\ref{eq:estep}) and are
shown in Fig.\ref{fig:estepp} as a function of $p$. Typically
terraces with $p\geq 6$ are wide enough to get the asymptotic
value, i.e., the isolated step energy with a numerical accuracy of
$\simeq 10^{-3}$eV. The values of the isolated step
energies for Rh, Pd and Cu are given in Table \ref{tab:energ_step} for the four
families of vicinal surfaces listed in Table \ref{tab:miller_indices}.

\begin{table}
\begin{center}
\begin{tabular}{c|c|c|c|c|c|c|c|c}
   & Vicinal surface            &    \multicolumn{7}{c}{Step energy $E_{step}$ (eV/atom)} \\
\hline
   &  $p \rightarrow \infty$     & TB & \multicolumn{6}{c}{EPP} \\
\cline{4-9}
   &                            &    &    & TB & Vitos & Methfessel & Eichler & Galanakis\\
\hline

Rh &  $p(111)\times(100)$       & 0.638 & $2V_1+4V_3$  & 0.657 & 0.583 & 0.520 & 0.650 & 0.670\\
   &  $p(111)\times(\bar{1}11)$ & 0.645 & $2V_1+4V_3$  & 0.657 & 0.583 & 0.520 & 0.650 & 0.670\\
   &  $p(100)\times(111)$       & 0.393 & $V_1+2V_2$   & 0.407 & 0.288 & 0.265 & 0.295 & 0.285\\
   &  $p(100)\times(010)$       & 0.747 & $2V_1+2V_2$  & 0.738 & 0.550 & 0.480 & 0.580 & 0.596\\
\hline
Pd &  $p(111)\times(100)$       & 0.425 & $2V_1+4V_3$  & 0.429 & 0.460 & 0.423 &       & 0.500\\
   &  $p(111)\times(\bar{1}11)$ & 0.432 & $2V_1+4V_3$  & 0.429 & 0.460 & 0.423 &       & 0.500\\
   &  $p(100)\times(111)$       & 0.289 & $V_1+2V_2$   & 0.295 & 0.106 & 0.222 &       & 0.298\\
   &  $p(100)\times(010)$       & 0.536 & $2V_1+2V_2$  & 0.533 & 0.265 & 0.427 &       & 0.548\\
\hline
Cu &  $p(111)\times(100)$       & 0.348 & $2V_1+4V_3$  & 0.347 & 0.380 &  &   & 0.426\\
   &  $p(111)\times(\bar{1}11)$ & 0.345 & $2V_1+4V_3$  & 0.347 & 0.380 &  &   & 0.426\\
   &  $p(100)\times(111)$       & 0.191 & $V_1+2V_2$   & 0.192 & 0.200 &  &   & 0.241\\
   &  $p(100)\times(010)$       & 0.352 & $2V_1+2V_2$  & 0.359 & 0.363 &  &   & 0.456\\
\end{tabular}
\caption{Step energies for various vicinal geometries.
Several types of results are presented: the full tight-binding (TB)
calculation and calculations based on effective pair potentials $V_1, V_2, V_3$
(EPP) fitted on the (111), (100), and (110) surface energies obtained
from various methods: tight-binding, and {\sl ab-initio} methods
(Vitos {\sl et al.} \protect\cite{Vitos99}, Methfessel {\sl et al.}
\protect\cite{Methfessel92} and Eichler {\sl et al.} \protect\cite{Eichler96})
and Galanakis {\sl et al.} \protect\cite{Galanakis02}}
\label{tab:energ_step}
 \end{center}
 \end{table}
 
 \begin{table}
\begin{center} 
\begin{tabular}{|c|c|c|c|c|c|}
(terrace)$\times$(ledge) &   \textbf{Rh} & \textbf{Pd} &\multicolumn{3}{c}{ \textbf{Cu}} \\
\hline
            &  &  &\textbf{ This work} & {\bf Other calculations}  & \textbf{Experiments} \\
\hline
$(111)\times(100)$ &   0.339 &  0.249   &  0.143  & 0.092 \cite{Feibelman99}  & 0.113 $\pm$ 0.007  \cite{Giesen01}\\
\hline
$(111)\times(\bar{1}11)$ &   0.329 &    0.242 &  0.148  &  0.117   \cite{Feibelman99}    & 0.121$\pm$0.007 \cite{Giesen01}\\
\hline
$(100)\times(111)$ &   0.349 &   0.247  &  0.146  &  0.139 \cite{Liu93}   &  0.123\cite{Barbier96} 0.129$\pm$0.009 \cite{Giesen01}\\
\hline
 $V_{1}$    & 0.332   &   0.238  & 0.166   &       &
\end{tabular}
\caption{ Kink energies for various steps with closed-packed edges in Rh, Pd and Cu (in eV) }
\label{tab:energ_kink}
\end{center}
\end{table}

Let us compare our results with those deduced from the effective
pair potential model proposed by Vitos et al.\cite{Vitos99}. In
this model the energy of a bulk atom is written:

\begin{equation}
E_{bulk}= -\sum_{R_J<R_c} Z^J_bV_J
\end{equation}

\noindent where $Z^J_b$ is the number of $J^{th}$ neighbors
at the distance $R_J$ for a
bulk atom and $R_c$ the cut-off radius of interactions, and the
surface energy (per surface atom) is:

\begin{equation}
E({\bf n_0},p) = \sum_{R_J<R_c}n^J_SV_J ,
\end{equation}

\noindent $n^J_S$ is the total number of $J^{th}$ neighbors (per
surface atom) suppressed by the surface. The step energies are
then given by:

\begin{equation}
E_{step}({\bf n_0},p) = \sum_{R_J<R_c}n^J_{step}({\bf n_0},p)V_J
\end{equation}

\noindent with:

\begin{equation}
n^J_{step}({\bf n_0},p) =n^J_S({\bf n_0},p) - (p-1-f)n^J_S({\bf n_0},\infty)
\end{equation}

\noindent where $n^J_S({\bf n_0},p)$ and $n^J_S({\bf n_0},\infty)$ are, respectively,
the total number of neighbors (per surface atom) in the $J^{th}$ coordination shell
suppressed by the vicinal and flat surfaces. In Vitos et al. work
\cite{Vitos99} the effective pair potentials are limited to first,
second and third neighbors ($V_1, V_2, V_3$) and their numerical
values are derived from the (111), (100) and (110) surface
energies. These surface energies are calculated using ab-initio
codes, the surface relaxation being neglected. Note that, due to
the short range of the pair potentials, the numbers
$n^J_{step}({\bf n_0},p)$ ($J\leq 3$) are constant as soon as $p$ exceeds a
given value $p_{\infty}$ which is actually very small, i.e., most
often $p_{\infty}=2$. As a consequence this model ignores
step-step interactions. Our calculations allow to check the
validity of this approach by using the pair potentials drawn from
the surface energies obtained with the same parametrised TB
hamiltonian. The expression for the step energies are given in
Table 3 and their numerical values are very close to the values
obtained from the previous method. Thus the method proposed by
Vitos et al. is quite valid to derive a good estimate of the step
energies when the low index surface energies and the step energy
calculated from (\ref{eq:estep}) are computed in the same
manner. Indeed, we have compared our results with those obtained
with the approach of Vitos et al. by using other data sets for the
surface energies
\cite{Vitos99,Methfessel92,Eichler96,Galanakis02}. It is seen in
Table 3 that the agreement is reasonable save for Pd
$p(100)\times(111)$ using the surface energy data set of Vitos et
al. Actually, with the latter data $V_2$ is negative, while it is
positive in the other calculations.

Finally note that in the effective pair potential model the
step energy of vicinal surfaces with (111) terraces is the same
for both ledge orientations $(100)$, i.e., type A step and
$(\bar{1}11)$, i.e., type B step (see Table 3) while in the full
TB calculation step A is slightly energetically favoured for Rh
and Pd, the reverse being found for Cu. This has some consequences
on the equilibrium shape of large adislands in homoepitaxy as will be
shown below.

\begin{figure}[!fht]
\begin{center}
  \ifpdf
  \includegraphics*[width=10cm]{figure3.pdf}
  \else
  \includegraphics*[width=10cm]{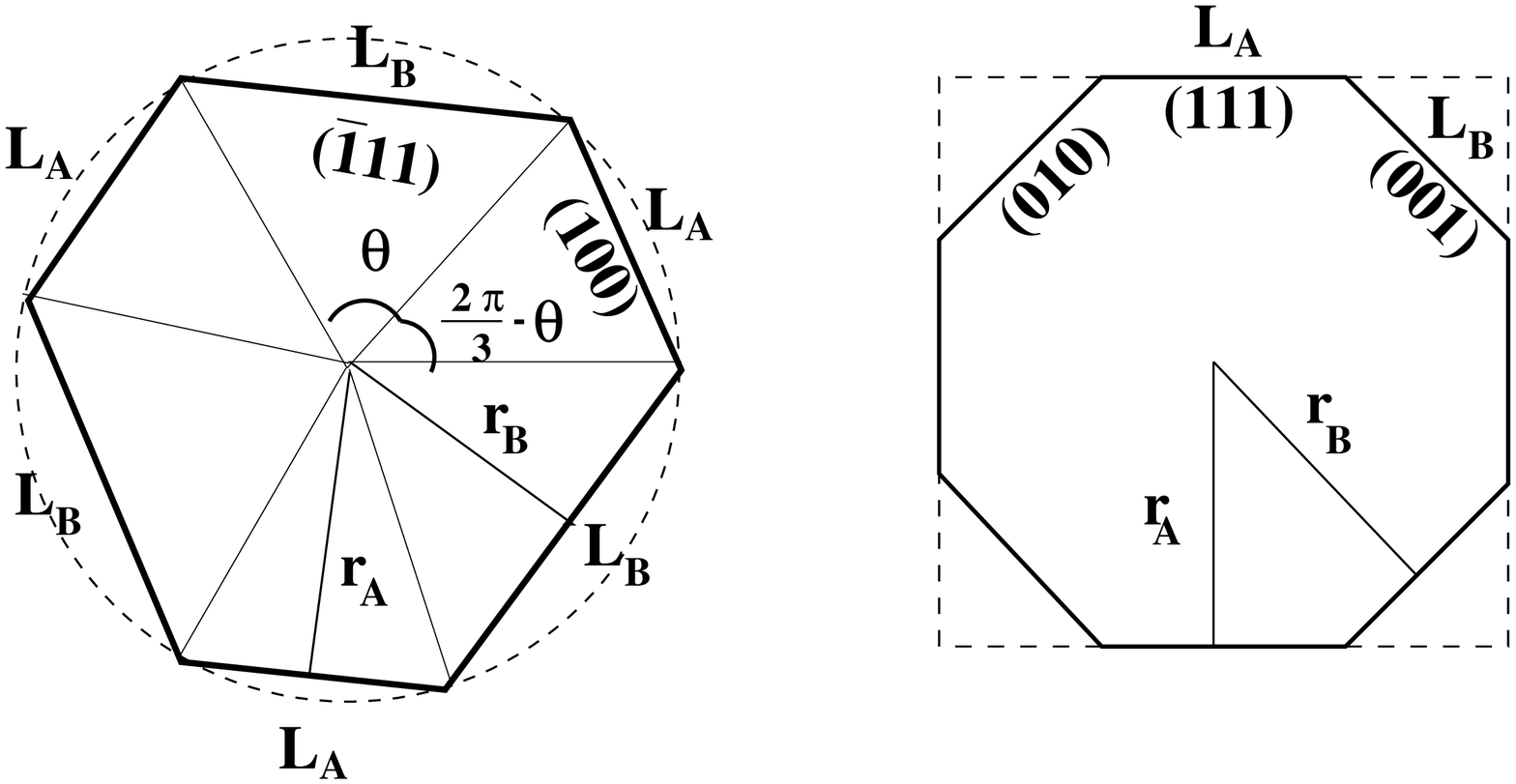}
  \fi
\caption{The equilibrium shapes of islands on (111) (a) and
(100) (b) FCC surfaces. The orientations of the microfacets are indicated.}
\label{fig:island}	  
\end{center}
\end{figure}

Let us now compare our results with experimental data. The ratio
of step energies and their absolute values can be directly
determined, by means of STM, from the observation, as a function
of temperature, of the equilibrium shape of 2D adislands in
homoepitaxy on a surface. On the (111) surface the adislands should show three-fold
symmetry. Consequently their shape is hexagon-like with
alternating A- and B-type edge segments (see
Fig.\ref{fig:island}). When the two step energies are equal, the A
and B segments have equal lengths and the hexagon is regular.
Otherwise the ratio $L_A/L_B$ of the lengths of A and B segments
can be derived from $\beta_A/\beta_B$ by applying the Wulff
theorem. Experimental results are only available
for Cu. The most recent experiments by Giesen et
al.\cite{Giesen01} give an average step energy for the two kinds
of steps on Cu(111) equal to $0.27\pm0.03$eV per step atom, the
energy of step A being measurably ($1.1\pm0.7\%$) larger than that
of a B step. Other published results \cite{Schlosser99,Icking99}
on the average step energy lie between 0.22 and 0.31eV. Our full
TB calculations (see Table 3) is in very good agreement with
experiments for the ratio $\beta_A/\beta_B$. However they seem to
slightly overestimate the average step energy. On the (100)
surface the adislands should show a four-fold symmetry. Accordingly, the most
simple polygonal shapes are a perfect square with (111) type
ledges or a square with broken corners with both (111) and (010)
type ledges (resp. A and B, see Fig.\ref{fig:island}). It can be
easily shown \cite{Raouafi102} that when
$E^A_{step}/E^B_{step}\leq 1/2$ the equilibrium shape is a perfect
square and a square with broken corners otherwise. The
experimental results of Giesen et al.\cite{Giesen01} show
unambiguously square adislands with broken corners from which they
deduce the ratio $E^A_{step}/E^B_{step}\simeq 0.57$ to be compared
to our full TB calculations that gives 0.54. The calculated energy
of the step $(100)\times(111)$ is 0.191eV, also in good agreement with
the experimental results ($0.22\pm0.02$eV).

\subsection{Step-step electronic interactions}

Let us now discuss step-step interactions. There exists several
types of interactions between steps. The most studied is the
so-called elastic interaction due to the deformation fields around
each step which interact repulsively. This elastic interaction
gives rise in the continuum elasticity limit to an energy term
varying at large inter-step distance as $1/d^2$ where $d$ is the
distance between two steps \cite{Marchenko81}. However, as we
will see below (Sect.4.2), when trying to fit results derived from
empirical potentials on relaxed surfaces, it appears that for
smaller $d$ ($d\leq$ 6 inter-row spacings) the behavior of
$E_{step}(d)$ deviates significantly from this law
\cite{Tian93,Wolf92,Najafabadi94,Hecquet96}. Furthermore
meandering steps cannot cross each other. This gives rise to an
entropic repulsive interaction varying as $1/d^2$ at large $d$
\cite{Jayaprakash84}. Charge transfers in the vicinity of the
steps produce a dipole-dipole interaction (repulsive or
attractive) varying also as $1/d^2$.

Finally  oscillatory
electronic interactions of the Friedel type (i.e., arising from
the interference between electron density oscillations around steps which have been
visualized by STM \cite{Hasegawa93}) should also be present
similarly to those existing between
chemisorbed atoms or defects \cite{Einstein73,Yaniv81}
but they have attracted little attention, at least up to now.
Such interactions have been invoked by Frohn {\sl et al.} \cite{Frohn91}
to explain their STM observations on Cu(11n). They have 
been introduced theoretically for cubium with a TB $s$ band by Redfield and
Zangwill \cite{Redfield92} and discussed in a phenomenogical manner by Pai
{\sl et al.} \cite{Pai94}. A calculation of these interactions for
vicinal surfaces of W(110) has also been carried out using a
modified fourth moment approximation to TB theory for a
pure $d$ band \cite{Wei96}.
However, until recently there were no detailed electronic structure
calculations on this subject, except one preliminary attempt with a
tight-binding scheme for FCC transition metals with a pure $d$
band showing that these oscillatory interactions do exist
\cite{Papadia96}. General trends were put forward but the
results were not quantitative due to the role played by $sp$
electrons in the total energy, which is significant in FCC
transition metals. The $spd$ TB model described above avoids this
approximation. Three main features can be extracted from
Fig.\ref{fig:estepp} (i) the step-step electronic interaction has a damped
behavior which is most often oscillatory, (ii) the amplitude of
the oscillations can be as large as some $10^{-2}$eV for small
values of $p$ and remains of the order of some $10^{-3}$eV when
$p\geq 5$ in the studied domain of $p$, (iii) the shape of the
oscillations is quite stable for two neighboring elements in the
periodic table (Rhodium and Palladium) but it is dependent on the
orientation of the steps.

Let us now compare our results with related works. The electronic
step-step interaction energies are of the same order of magnitude
as the full step-step interactions derived from experiments by using
an analysis of terrace width distributions which most often
assumes purely repulsive interactions varying as $1/d^2$. This
suggests that, as already mentioned \cite{Pai94}, this type of
interactions should be included in the treatment of experimental
data. Nevertheless, it remains difficult to fit our results by an
analytical expression and extrapolate an asymptotic behavior to
compare with elastic interactions.

Unfortunately they are only a few experimental data in the domain
of small terrace widths. However, an anomalous behavior of the
terrace width distribution at low temperatures for Cu
$p(100)\times(111)$ has been observed by Frohn et
al.\cite{Frohn91} which could be interpreted as due to a repulsive
interaction when $d\simeq 1-2$ (in units of the nearest neighbor
distance) but attractive (or oscillatory) when $d\simeq 3-5$. This
is quite consistent with our results (Fig.\ref{fig:estepp}).

Finally kink formation energies have been calculated within the
same TB method \cite{Raouafi102} and the geometry suggested by Feibelman
\cite{Feibelman99}. The results are given in Table \ref{tab:energ_step} and compare
favourably with existing experiments and other calculations \cite{Feibelman99,Liu93}.

\section{Vibrational properties of Cu vicinal surfaces}

The presence of steps on a surface modifies the electronic
structure (see Sect.3) as well as the vibrational states,
compared to the flat surface with the same orientation as the
terraces. In the following we will use an empirical potential to
investigate the vibrational states of vicinal surfaces of copper
and deduce the contribution of phonons to the free energy of
steps.

\subsection{The empirical potential.}

The empirical potential used to describe the interatomic interactions of a set
of atoms located at $\bf R_i$ is of the form:

\begin{eqnarray}
V({\bf R_1},...{\bf R_i},...) &=& A \sum_{i,j, j\ne i} (R_0/R_{ij})^p f_c(R_{ij}) \nonumber \\
   &-& \xi \sum_i \Big[\sum_{j \ne i} \exp[-2q(R_{ij}/R_0-1)] f_c(R_{ij}) \Big]^{\alpha}
\label{potential}
\end{eqnarray}

\noindent where $R_{ij}$ is the distance between atoms $i$ and
$j$, $R_0$ is a reference distance that we take equal to the bulk
nearest neighbor spacing, $f_c(R)=1/(1+exp[(R-R_c)/\Delta])$ is
a smooth cut-off function with a cut-off radius $R_c$ and $\alpha$
is an exponent set equal to $2/3$.

The parameters $A$, $\xi$, $p$ and $q$ are fitted to the
experimental values of the cohesive
energy $E_c$ ($E_c=-3.5eV/at$) and of the three elastic constants, i.e.,
the bulk modulus ($B=10.470eV/at$) and the two shear moduli $C$ and $C'$
($C=6.046eV/at, C'=1.917eV/at$). The
equilibrium equation at $R_0=2.5526${\AA}  gives a relation between
the four parameters.

We have determined by a least mean square fit the sets of
parameters obtained with different radii $R_c$ for which
interactions are cut off beyond first, second, third and fourth
neighbors. For each set of parameters we have compared the fitted
values of $E_c, B, C, C'$, the surface relaxation of low index
surfaces and the bulk phonon spectra to experiments. Let us
mention that we have also tried other sets of exponents for
$\alpha$ but  the choice of $\alpha=2/3$ was greatly improving the
surface energies compared to ab-initio data \cite{Barreteau02}.

The best set of parameters is obtained for  a cut-off radius
$R_c=4.02${\AA} between second and third neighbors and the
corresponding parameters are: A=0.206eV, $\xi$=1.102eV, p=7.206,
q=2.220. Indeed with this potential
(hereafter referred to as $P_2$) the fit of $E_c, B, C$ and
$C'$ is excellent (better than 1meV per atom). When the potential
includes only first nearest neighbors the shear moduli are not
well reproduced and, in particular, $C$ is about 25\% smaller than
the experimental value. The inclusion of third and fourth
neighbors has a smaller influence on the elastic constants but the
general tendency is an underestimation of the inward surface
relaxation as one increases the cut-off radius beyond the second
neighbors. We must emphasize that surface relaxation is important
to get the local modifications of force constants correctly.

\begin{figure}[!fht]
\begin{center}
  \ifpdf
  \includegraphics*[width=10cm]{figure4.pdf}
  \else
  \includegraphics*[width=10cm]{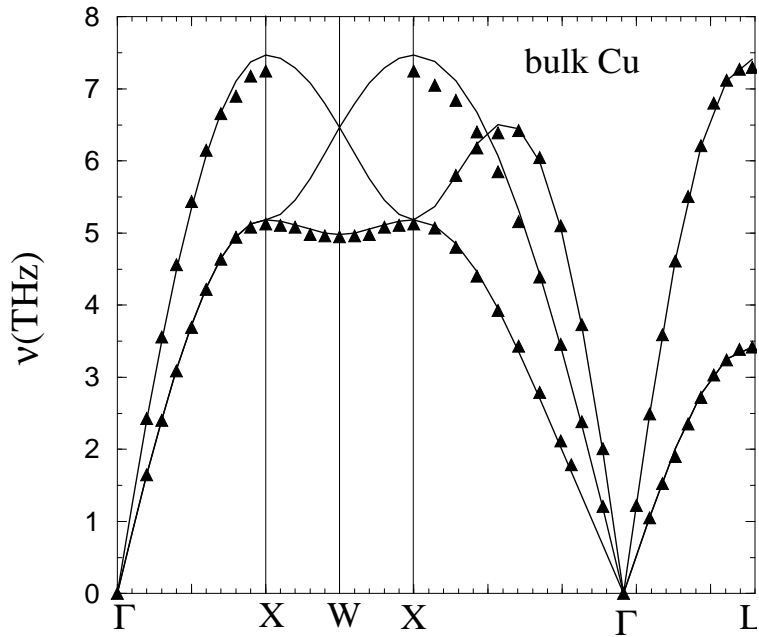}
  \fi
\caption{Phonon dispersion curves of bulk copper. The full lines correspond to
the calculated dispersion curves and  the triangles  to the
phonon frequencies measured from neutron inelastic-scattering
experiments (Ref. \protect\cite{Nicklow67}). Each segment along the path
$\Gamma (\Delta) X (Z) W (Z) X (\Sigma) \Gamma (\Lambda) L$ is proportional
to its length in reciprocal space.}
\label{fig:bulk_phonons}	  
\end{center}
\end{figure}

To obtain the phonon dispersion curves in the harmonic
approximation the dynamical matrix is calculated from the
analytical expression of the potential (\ref{potential}) and
diagonalized for wave vectors ${\bf k}$ following symmetry lines
in the BBZ. The calculated bulk dispersion curves are presented in
Fig \ref{fig:bulk_phonons}. The agreement with experiment
\cite{Nicklow67} is excellent. Apart from the top of the spectrum
at points X and L where the deviation between calculated and
experimental frequencies is around 0.2THz, everywhere else the
deviation is less than 0.1THz. Note also that the shallow minimum
at W in the lowest frequency band along XWX is reproduced only
when $R_c$ is chosen between second and third nearest neighbors.

We have also calculated the surface projected band structure of
phonons for the three low index surfaces $(111)$, $(100)$ and
$(110)$ using the usual slab geometry. Our results were compared
with available experimental EELS and IHAS data. The agreement is
excellent for the three surfaces \cite{Barreteau02}. Low frequency
as well as high frequency surface localized modes are reproduced
with a surprising accuracy which is a good check of the
transferability of the potential since surface modes are extremely
sensitive to local modifications of the force constants due to the
surface relaxation.

\subsection{Atomic relaxation and elastic step-step interactions.}

The first task is the determination of the equilibrium atomic
structure which is obtained by a standard conjugate gradient
method. A common feature for most metallic vicinal surfaces is
that all atoms, save at the inner edge, relax inwards,i.e. ,
towards the bulk similarly to low index surfaces. However the
direction of relaxation changes with the position $p$ of the
atomic row on the terrace and one can identify a vortex-like
structure described in a recent paper by Pr\'evot {\sl et al}.
\cite{Prevot02} The outer edge step atom (SC: step chain) always
shows the largest inward relaxation, therefore the distance
between the outer edge atom and its first nearest neighbor having
the bulk coordination (BNN) exhibits the largest contraction
compared to the bulk equilibrium nearest neighbor distance. As
will be seen later the shortening of SC-BNN bonds produces a
stiffening of the associated force constant.The inner edge atoms,
contrary to the other terrace atoms, relax outwards. Another
common feature to all metallic vicinal surfaces is the profile of
the multilayer relaxation, defined as the ratio of the distance
between two adjacent atomic planes parallel to the vicinal surface with
respect to the corresponding bulk inter-layer spacing. The
multilayer relaxation always shows a damped oscillatory behavior
with a period of oscillation equal to the vicinality $p$ of the
surface \cite{Barreteau02,Tian93}.

\begin{figure}[!fht]
\begin{center}
  \ifpdf
  \includegraphics*[width=15cm]{figure5.pdf}
  \else
  \includegraphics*[width=15cm]{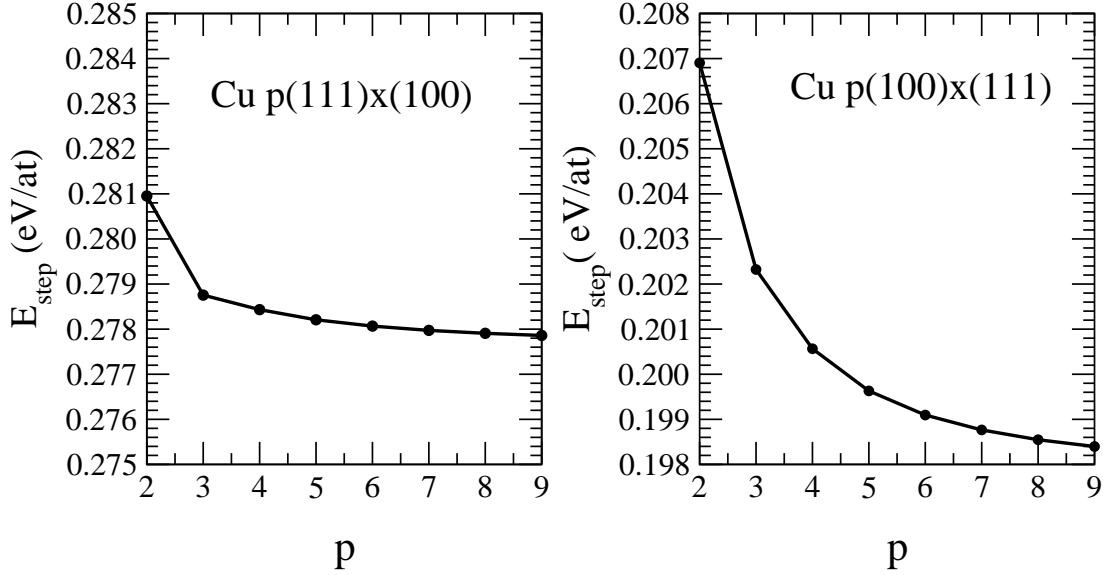}
  \fi
\caption{Variation of the step energy per step atom of the $p(100)\times(111)$
and $p(111)\times(100)$ vicinal surfaces as a function of the terrace width $p$.
The geometry has been fully relaxed.}
\label{fig:elastic_step_step_interaction}	  
\end{center}
\end{figure}

The equilibrium structure of a vicinal surface with a given terrace width $p$
 being known, it is straightforward to calculate the corresponding step energy
 per atom $E_{step}({\bf n_0},p)$ using equation (\ref{eq:estep}).
 $E_{step}({\bf n_0},p)$ is varying with the terrace width $p$ as a result of step-step
interactions. The step energy is obtained in the limit
$p\rightarrow\infty$. In Fig.
\ref{fig:elastic_step_step_interaction} we have presented the step
energy per step atom for the vicinal $p(100)\times(111)$  and
$p(111)\times(100)$ surfaces for $p$ ranging from 2 to 9.
$E_{step}({\bf n_0},p)$ is strictly decreasing when $p$ increases as
expected from a calculation based on a semi-empirical potential ,
since no oscillatory electronic effects are taken into account
\cite{Raouafi102}. This variation is the result of purely elastic
step-step interactions which are known to be repulsive from
elasticity theory and, as mentioned previously, this term should
vary at large inter-step distances as $1/d^2$ where $d$ is the
distance between two adjacent steps \cite{Marchenko81}. In order
to compare the present results with the prediction of the
elasticity theory we have fitted the step energy per step atom
$E_{step}$ as a function of $d=(p-1+f)a_0$ where $a_0$ is
the distance between two adjacent atomic rows on the terrace plane
($ 6 \le p \le 100$) , with an expression of the form
$A_0+A_2/d^2+A_3/d^3$. In the case of Cu $p(100)\times(111)$
surfaces, $f=0.5$ and $a_0=R_0$. The result is $A_0=0.198eV$,
$A_2=0.322eV$\AA$^2$ and $A_3=-0.955eV$\AA$^3$. $A_0$ is the
asymptotic value giving the step energy of an isolated step
and is in very good agreement with the value given by TB calculations
(see Table 3). It is
interesting to note that the coefficients $A_2$ and $A_3$ have
opposite signs similarly to what was found on $p(100)\times(010)$
surfaces of Ni and Au in a previous atomistic study
\cite{Srolovitz96}. Although $A_3$ is non-zero, its contribution
becomes negligible for $p$ larger than 10.

\subsection{Projected phonon band structure of vicinal surfaces.}

The vibrational spectra of  various vicinal surfaces have been
presented in a recent paper \cite{Barreteau02}. Thus, the
most important features will be illustrated here on two specific cases
for which detailed experimental IHAS \cite{Witte95} and EELS
\cite{Kara00} data are available for phonons propagating parallel
and perpendicular to the step edge on the $(211)$ and $(511)$
surfaces.

There are several common features on the surface projected band
structure of vicinal surfaces. First the most striking feature is the
disappearance of almost all gaps in any direction of ${\bf
k_{//}}$ space,  for the same reason as already explained in the
electronic structure section. Second, the localized modes
propagating perpendicular to the terraces have a clear back-folded
structure. This back-folding leads to optical modes, their number
increasing with the terrace width. Third, the localized modes
propagating parallel to the step edge show strong similarity with
the corresponding modes of the low index surface with the same
terrace orientation. Finally some resonant or localized modes in
the vicinity of the steps also appear. In particular, a very
specific mode is present on almost all dispersion curves at the
very top of the band edge. This state is purely localized on the
BNN atoms and is closely related to the stiffening of the force
constant mentioned above.

\begin{figure}[!fht]
\begin{center}
  \ifpdf
  \includegraphics*[width=10cm]{figure6a.pdf}
  
  \includegraphics*[width=10cm]{figure6b.pdf}
  \else
  \includegraphics*[width=10cm]{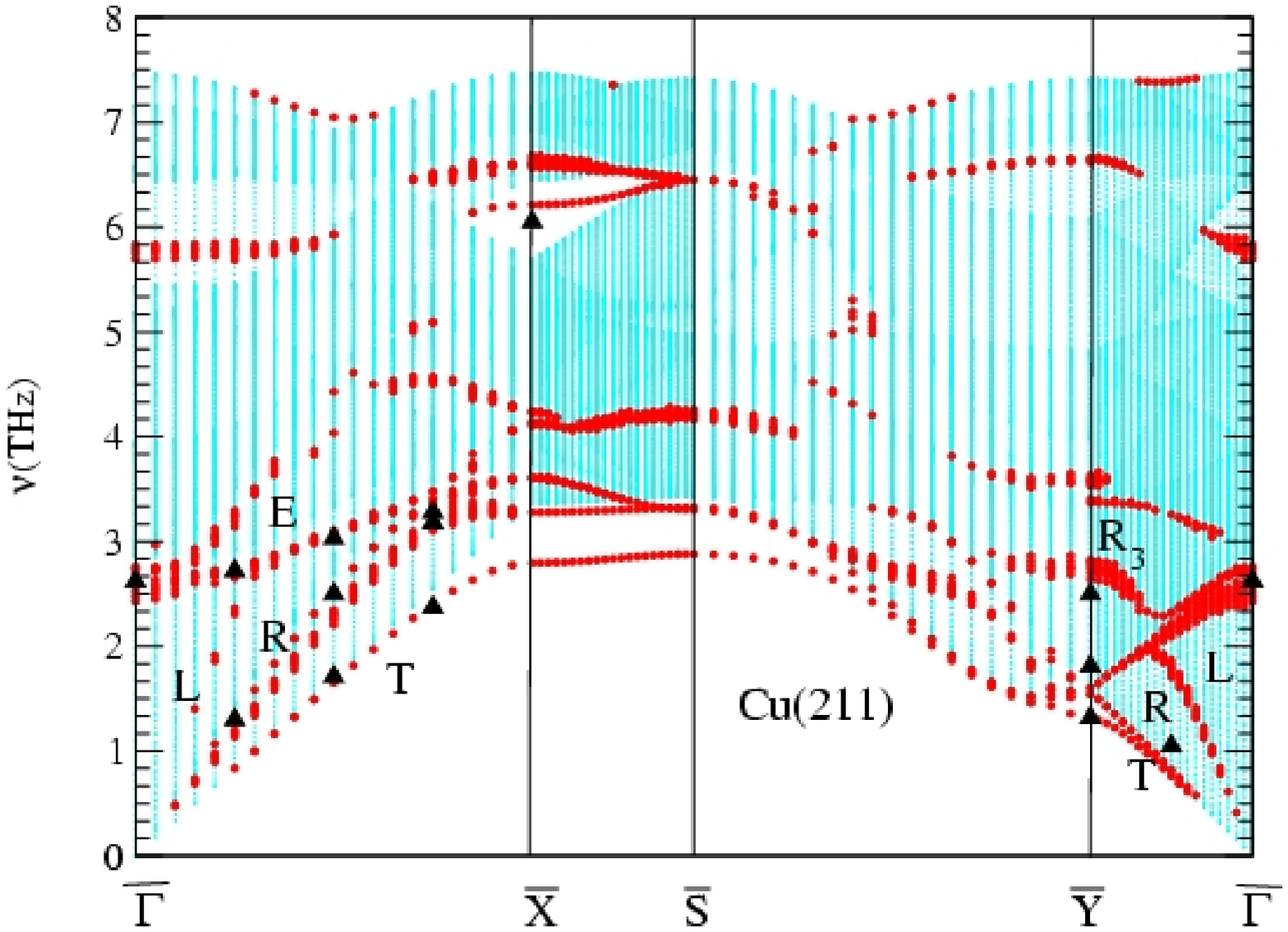}
  
  \includegraphics*[width=10cm]{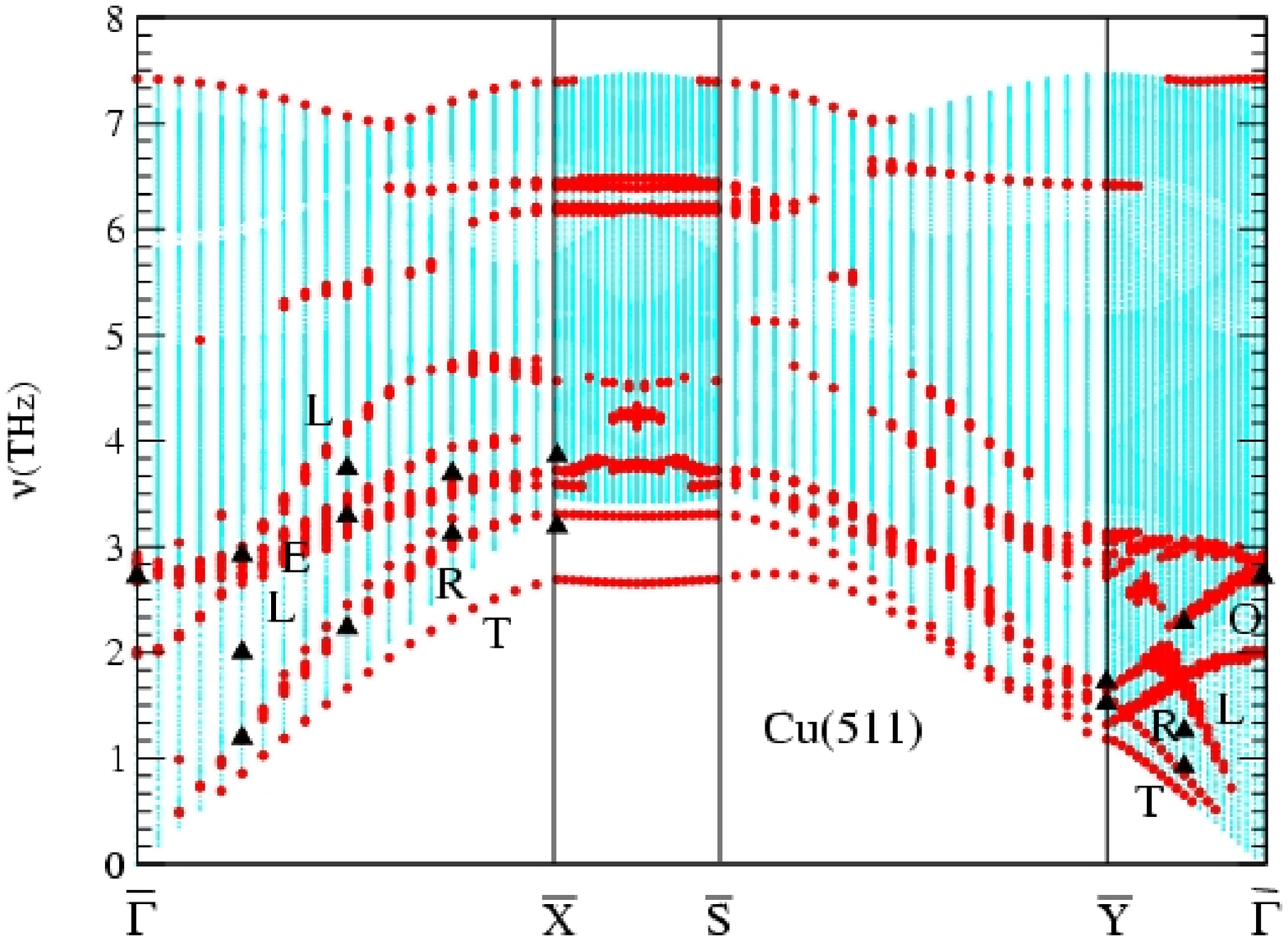}  
  \fi
\caption{Frequency spectrum of phonons for the (211) (a) and (511) (b) surfaces of Cu as a function
of {\bf k$_{//}$} along a given path  in the surface
Brillouin zone.  Each segment is proportional to its
length in the reciprocal space. Bulk states are represented by small dots.
Localized and resonant states are denoted by heavy dots.
The prominent surface features are
denoted with the same notations as in Ref.\protect\cite{Witte95} and the
localization criterion is 6\% on the first four layers. Experimental
points are taken from Ref.\protect\cite{Witte95} except for the highest
frequency at $\bar X$ which corresponds to the EELS data of Ref.\protect\cite{Kara00}.}
\label{fig:phonons_211_511}	  
\end{center}
\end{figure}

These general features are seen on the surface projected phonon
band structure of the $(211)$ and $(511)$ surfaces, i.e., with Lang
et al. notations $3(111)\times(100)$ and $3(100)\times(111)$.
Indeed, in Fig.\ref{fig:phonons_211_511}, the only small noticeable gap
is around the $\bar X$ point for the $(211)$ surface, whereas all
gaps have disappeared for the $(511)$ surface. Let us consider the
$\bar \Gamma \bar X$ and $\bar \Gamma \bar Y$ directions
corresponding to directions of propagation parallel and
perpendicular to the step edges, respectively. Along the $\bar
\Gamma \bar X$ direction of the $(211)$ surface, the most
prominent surface features are a transverse mode (T) horizontally
polarized, the sagittal Rayleigh mode (R) and a step localized
mode (E). In addition another localized mode is found around the
middle of the gap at $\bar X$. The same type of modes are also
found on the $(511)$ surface. Moreover there is a weakly localized
longitudinal mode (L) which, actually, is also present on the
(211) surface but is even less localized. All these results are in
very good agreement with IHAS and EELS data. Modes propagating
perpendicularly to the step (along $\bar \Gamma \bar Y$) can be
qualitatively described as resulting from a ``back-folding"
\cite{Armand83} of the Rayleigh and transverse modes. Note that
the mode localized on the BNN atom is particularly visible on the
$(511)$ surface for which the $z$ component of the force constant
between the BNN and step edge atom is $44\%$ larger than the bulk
one \cite{Barreteau02}.

\subsection{Vibrational free energy of steps.}

\begin{figure}[!fht]
\begin{center}
  \ifpdf
  \includegraphics*[width=15cm]{figure7.pdf}
  \else
  \includegraphics*[width=15cm]{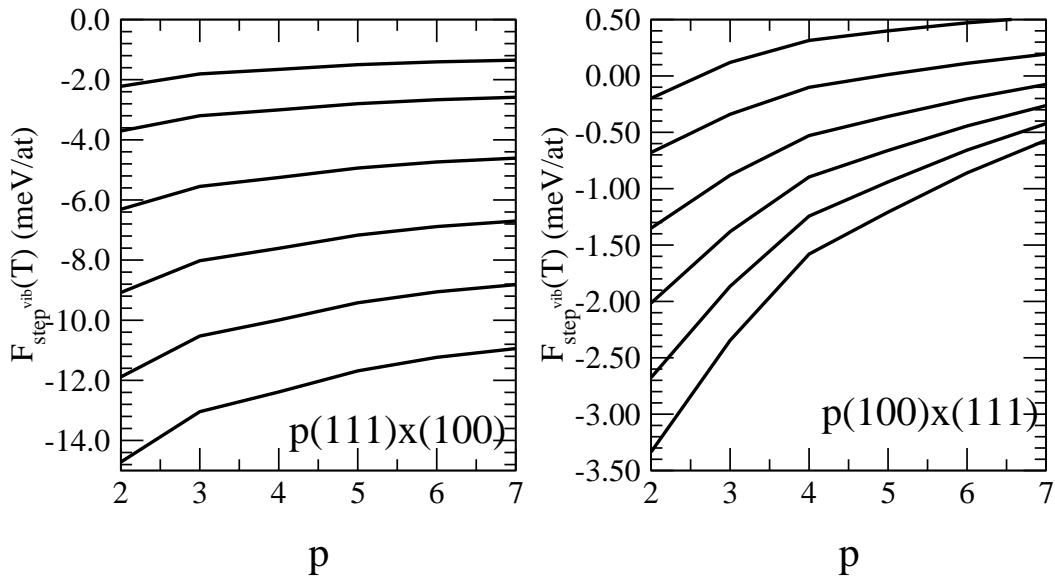}
  \fi
\caption{The contribution of vibrations to the step free energy for
$p(100)\times(111)$ and $p(111)\times(100)$ vicinal surfaces as a function
of $p$ for given temperatures.}
\label{fig:estep_phonons}	  
\end{center}
\end{figure}

The contribution of vibrations to the free energy of a system
which has a total density of frequencies $n(\nu)$ is given by:

\begin{equation}
F^{vib}(T)=k_BT\int_{0}^{\infty}\ln(2\sinh\frac{h\nu}{2k_BT})n(\nu)d\nu
\end{equation}

\noindent where $k_B$ is the Boltzmann constant. 
In the case of 2D periodicity, the integral over
the frequency is carried out by summing over special $\bf k_{//}$
points belonging to the irreducible SBZ. From the vibrational free
energy of vicinal surfaces, low index surfaces and bulk, the
vibrational free energy of steps (per step atom)
$F_{step}^{vib}(T)$ can be derived at any temperature using an
equation similar to (\ref{eq:estep}). We have calculated
$F_{step}^{vib}(T)$ for the $p(100)\times(111)$ and
$p(111)\times(100)$ vicinal surfaces of increasing terrace widths,
for temperatures ranging from $0$ to $500$K. The step vibrational
free energy of a given vicinal surface is of the order of a few
meV and decreases with temperature, reaching a linear regime for
$T$ larger than $100$K when the entropy contribution becomes the
leading term (see ref. \cite{Mcdsbook}). More interestingly
$F_{step}^{vib}(T)$ can be plotted for a given temperature, as a
function of the terrace width as shown in Fig.
\ref{fig:estep_phonons}.

It appears that $F_{step}^{vib}(T)$
decreases in absolute value when the terrace width increases,
i.e., phonons produce attractive step-step interactions.
Furthermore the absolute value of these attractive step-step
interactions increases with temperature. The possibility of
interactions between surface defects mediated by phonons has
already been investigated by Cunningham {\sl et
al.}\cite{Cunningham73} who have derived the phonon contribution
to the free energy of interaction for an adatom pair on the (100)
face of cubium using the Montroll-Potts model and also found an
attractive interaction. Finally note that, even though the
vibrational contribution to the step energy is of the order of a
few meV, therefore quite negligible compared to the absolute value
of the step energy, its variation with the vicinality can be of
the same order of magnitude as the repulsive elastic one at least
in the range of small terrace widths ($p<10$).

\begin{figure}[!fht]
\begin{center}
  \ifpdf
  \includegraphics*[width=15cm]{figure8.pdf}
  \else
  \includegraphics*[width=15cm]{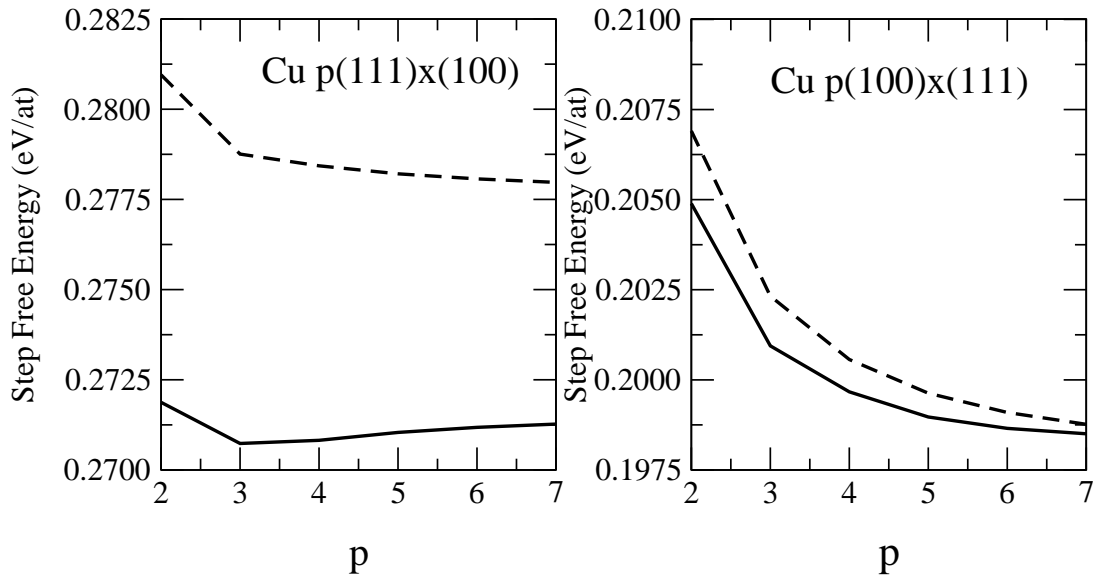}
  \fi
\caption{Variation of the step free energy at 300K as a function of $p$ without
(dashed line) and with (full line) the vibrational contribution.}
\label{fig:sum_step_energy}	  
\end{center}
\end{figure}

Fig.\ref{fig:sum_step_energy} shows the variation of the step
energy as a function of $p$ including both the elastic and
vibrational contributions at 300K. The correction to the isolated
step energy is small but phonons change the curvature  of the step
energy and, surprisingly, may even modify its sign. In particular
for the $p(111)\times(100)$ surfaces the step energy exhibits a
minimum at $p=3$, i.e. , the resulting step-step interactions are attractive.

In addition, it must be kept in mind that electronic effects
are far from  being negligible, at least for small terrace widths  (typically less than
10 atomic rows), and usually give rise to oscillatory interactions as shown
in the previous sections.

\section{Stability of vicinal surfaces with respect to faceting}

Vicinal surfaces are not always stable. Indeed, as seen in Sect.3,
their surface energies $\gamma$ are large and it might be
energetically favorable for the solid to expose to vacuum low
index facets with smaller surface energies per unit area, even if
the total surface area is increased by the transformation. This
phenomenon, called faceting, has been known for a long time
\cite{Rosenhain02}. The faceting condition implies the calculation
of the surface energy for any surface orientation, i.e., the
knowledge of the $\gamma$-plot. Herring \cite{Herring51} was the
first to propose a geometrical construction starting from the
$\gamma$-plot and predicting the occurrence of faceting. Then this
condition was recast in a much simpler way (see Sect.5.1). The
most simple methods for calculating the surface energies as a
function of the orientation range from the crudest empirical pair
potentials to various semi-empirical ones (EAM, EMT, SMA...),
including an N-body contribution, which have been set up in the
last two decades. Recently, the stability of vicinal surfaces with
respect to faceting was reexamined using the EMT potential
\cite{Frenken99}. It was shown that the total energy difference
between the vicinal and facetted surface is very small and,
surprisingly, it was found that all vicinal surfaces between the
$(100)$ and $(111)$ planes were unstable, at least at 0K, and that
the observed stability at room temperature arises from the entropy
contribution due to thermal vibrations. However, semi-empirical
potentials have a common drawback: they only depend on the
interatomic distances and not on the angular arrangement of atoms.
Although this latter effect is small in metals, it is not obvious
that it can be neglected in view of the tiny energy difference
involved in faceting. In Refs.\cite{Desjonqueres02,Raouafi202}, we
have revisited this problem at 0K and analyzed the answers given
by pair potentials, semi-empirical potentials and the TB
calculations presented above (Sect.3). Finally our study of
vibrational properties of vicinal surfaces (Sect.4) has enabled us
to investigate the effect of finite temperatures.

In the following we first recall the faceting condition. Then we
summarize our results concerning the possible faceting of the
vicinal surfaces that are spanned when going from the $(100)$ to
the $(111)$ plane. Other domains of orientations have been studied
in \cite{Raouafi202}.

\subsection{Faceting condition of an infinite surface}

Let us consider two low index surfaces $\Sigma_1$ and $\Sigma_2$
with normals {$\bf n_1$} and {$\bf n_2$}, respectively, which
intersect along a given row of atoms and the set of vicinal
surfaces with equidistant step edges which is spanned when
$\Sigma_1$ is rotated around the common atomic row towards
$\Sigma_2$. Let us take $\Sigma_1$ as the origin of angles and
denote $\theta_2$ the angle ({$\bf n_1$},{$\bf n_2$}). During this
rotation the surfaces vicinal to $\Sigma_1$ are first found and
the number of atomic rows $p_1$ (including the inner edge) on one
terrace decreases from $\infty$ to 2 (angle $\theta_c$). The
surface corresponding to $\theta_c$ can also be regarded as a
vicinal of $\Sigma_2$ with $p_2=2$. Then for $\theta_c \leq \theta
< \theta_2$ the surfaces vicinal to $\Sigma_2$ are scanned with
increasing terrace widths ($p_2 \geq 2$). An area $S$ of any of
these high index surfaces will transform into facets of normal
{$\bf n_1$ (area $S_1$) and normal {$\bf n_2$ (area $S_2$) while
keeping its average orientation when (Fig.\ref{fig:faceting})

\begin{equation}
\gamma S > \gamma_1 S_1 + \gamma_2 S_2 \label{eq:gammas}
\end{equation}

\begin{figure}[!fht]
\begin{center}
  \ifpdf
  \includegraphics*[width=15cm]{figure9.pdf}
  \else
  \includegraphics*[width=15cm]{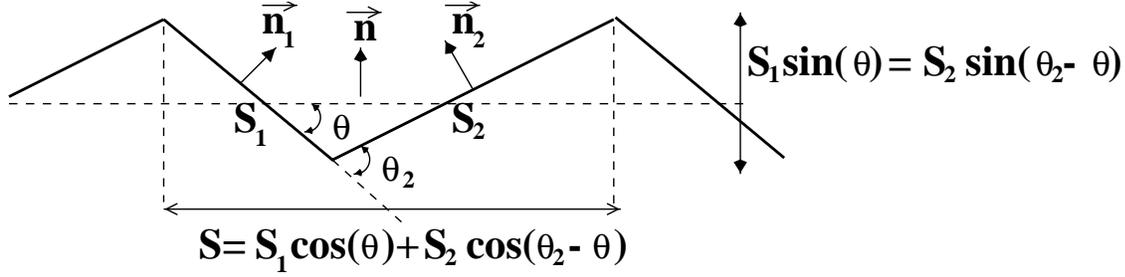}
  \fi
\caption{Faceting}
\label{fig:faceting}	  
\end{center}
\end{figure}

\noindent ($\gamma, \gamma_1$ and $\gamma_2$ being the surface
energies per unit area of the high index, $\Sigma_1$ and
$\Sigma_2$ surfaces, respectively) with the constraints:

\begin{eqnarray}
S=S_1 \cos \theta + S_2 \cos(\theta_2-\theta) \\
S_1 \sin \theta = S_2 \sin(\theta_2-\theta) \label{eq:constraints}
\end{eqnarray}

\noindent It is easily shown that the faceting condition can be
written

\begin{equation}
f(\eta) > (1 - \eta/\eta_2)f(0) + (\eta/\eta_2)f(\eta_2)
\label{eq:feta}
\end{equation}

\noindent with $\eta=\tan \theta$ and $f(\eta)=\gamma(\theta)/\cos
\theta$. This condition is equivalent to the Herring
construction \cite{Raouafi202}.

This inequality has a simple geometrical interpretation: the
vicinal surface corresponding to $\eta$ is unstable(stable) when
the point ($\eta,f(\eta)$) is above(below) the straight line D joining the
points ($0,f(0)$) and ($\eta_2,f(\eta_2)$) or, equivalently, the
sign of the deviation $\Delta f(\eta)$ from this straight line
determines the stability ($\Delta f(\eta) < 0$) or the instability
($\Delta f(\eta) > 0$) of the vicinal surface. With
straightforward geometrical considerations, it is easily shown
\cite{Raouafi202} that:

\begin{equation}
\Delta f({\bf n}) = [E_S({\bf n}) - (p_1 - 1)E_S({\bf n_1}) - (p_2
- 1)E_S({\bf n_2})]/A_0({\bf n}) \label{eq:deltaf}
\end{equation}

\noindent where $A_0({\bf n})$ is the projected area of the
surface unit cell A of the vicinal surface of orientation ${\bf
n}$ on $\Sigma_1$. This formula applies as well in the domain $0
\leq \theta \leq \theta_c$ with $p_2=2$, as when $\theta_c \leq
\theta \leq \theta_2$ with $p_1=2$. $E_S({\bf n})$ is the surface
energy (per atom) of the surface normal to ${\bf n}$. It is
interesting to note that the condition of instability of the
surface corresponding to $\eta_c$ (normal ${\bf n_c}$) is simply:

\begin{equation}
E_S({\bf n_c}) > E_S({\bf n_1}) + E_S({\bf n_2}) \label{eq:esc}
\end{equation}

\noindent we will see below that in many cases the sign of $\Delta
f(\eta_c)$ determines the stability for the whole range
$[0,\eta_2]$. It is clear that the sign of $\Delta f$ is
independent of the origin of angles, i.e., if $\Sigma_1$ is
referred by the angle $\theta_1$, since it is given by the sign of
the expression between the square brackets in Eq.(\ref{eq:deltaf})
which will be denoted as $\Delta E(p_1,p_2)$ in the following.

Let us denote $A_1$ ($A_2$) the area of the unit cell of
$\Sigma_1$ ($\Sigma_2$). It is straightforward to show that:

\begin{eqnarray}
\left\{\begin{array}{r c l r}
\Delta f(\eta) &=& \displaystyle \frac{\Delta E(p_1,2)}{A_2\sin \theta_2}\eta  &     0 \leq \eta \leq \eta_c \\
   &  &  & \\
\Delta f(\eta) &=& \displaystyle \frac{\Delta E(2,p_2)}{A_1}(1 - \eta/\eta_2)  &     \eta_c \leq \eta \leq \eta_2 \\
\end{array} \right.
\label{eq:deltafeta}
\end{eqnarray}

By using equation (\ref{eq:estep}), $\Delta E(p_1,2)$ can be
transformed into:

\begin{equation}
\Delta E(p_1,2)= E_{step}({\bf n}_1,p_1) - E_S({\bf n}_2)+
f_1E_S({\bf n}_1)
\end{equation}

\noindent (a similar equation can be written for $\Delta E(2,p_2)$
by interchanging the indices 1 and 2 in the right hand side of this
equation). As seen in Sect.3, the step energy varies with $p$ due
to step-step interactions and, consequently, $\Delta E(p_1,2)$ and
$\Delta E(2,p_2)$ depend on $\eta$. When these interactions are
neglected, these last two quantities are equal to their values at
$p_1=2$ and $p_2=2$, or $\eta=\eta_c$. Then from
(\ref{eq:deltafeta}) $\Delta f(\eta)$ has a
triangular shape (Fig.\ref{fig:triangle}). 

\begin{figure}[!fht]
\begin{center}
  \ifpdf
  \includegraphics*[width=10cm]{figure10.pdf}
  \else
  \includegraphics*[width=10cm]{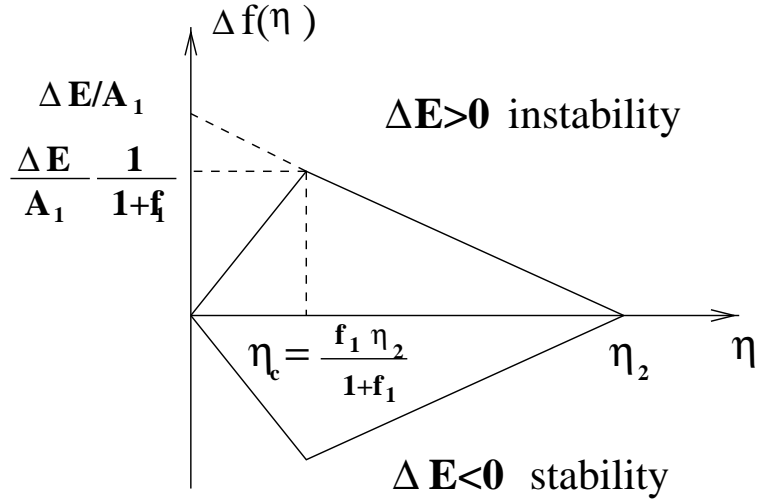}
  \fi
\caption{Behavior of $\Delta f(\eta)$ when there are no interactions between
steps. $\eta_c$ corresponds to $p_1=p_2=2$.}
\label{fig:triangle}	  
\end{center}
\end{figure}

Any
vicinal between $\Sigma_1$ and $\Sigma_2$ is unstable(stable)
relative to faceting when $\Delta E>0$($\Delta E<0$). As a
conclusion any deviation of $\Delta f(\eta)$ from the triangular
shape is the sign of the presence of interactions between steps.
If for a given orientation $\eta_0$ such that $0<\eta_0<\eta_c$,
$\Delta f(\eta_0)$ is above(below) the tangent to $\Delta f(\eta)$
at the origin, then the interactions between steps are
repulsive(attractive). The same conclusion holds for
$\eta_c<\eta_0<\eta_2$ but the tangent has to be taken at
$\eta=\eta_2$. In the domain that will be considered below and
defined by ${\bf n}_1(100)$ and ${\bf n}_2(111)$, i.e.,
$(100)-(111)$, $\eta$ varies from 0 to $\eta_2=\sqrt{2}$. When
$0<\eta \leq \eta_c (\eta_c=\sqrt{2}/3)$ the crystallographic
planes (2p-1,1,1) are spanned and correspond to the
$p(100)\times (111)$ surfaces 
and when $\eta_c \leq \eta < \eta_2$ ($\eta_2=\sqrt{2}$) the
crystallographic planes are (p+1,p-1,p-1) and the corresponding
vicinal surfaces are $p(111)\times (100)$ (see Table 1). Note that
for $\eta=\eta_c$ the Miller indices of the surface are (311).

\subsection{Stability of vicinal surfaces at 0K from
semi-empirical potentials}

Empirical potentials belonging to a very large class can be
written as a sum of contributions $E_i$ of each atom $i$ (referred
to the energy of a free atom, $E_i<0$) depending on its
environment of neighbors $j$ at the interatomic distance $R_{ij}$,
i.e.:

\begin{equation}
E=\sum_i E_i=\sum_i \Big\{\sum_{j\neq i} V(R_{ij}) + F \big(
\sum_{j\neq i} g(R_{ij}) \big)  \Big \} \label{eq:pot}
\end{equation}

\noindent $E$ is the total energy of the system at 0K neglecting
the zero point vibrational energy. In the following we set
$\rho_i=\sum_{j\neq i} g(R_{ij})$. The first term of
Eq.(\ref{eq:pot}) is thus pairwise while the second one (in which
$g$ is a positive function) has an N-body character. The functions
$V$ and $g$ are usually cut-off smoothly around a given radius
$R_c$. This class of potentials includes pair potentials
($F(\rho_i)=0$), potentials based on Effective Medium Theory
(EMT)\cite{Stoltze94,Jacobsen87}, Embedded Atom Model
(EAM)\cite{Daw84} and glue model \cite{Ercolessi88}, and
potentials derived from the tight-binding approximation in the
second moment approach ($F(\rho_i) \propto \sqrt{\rho_i}$)
\cite{Ducastelle70,Sutton84,Finnis84}or in which
($F(\rho_i) \propto \rho_i^{\alpha}$). This law
with $\alpha=2/3$ has been proposed to account for the effect of
higher order moments \cite{Guevara95} and has been actually found
for Cu when fitting the five parameters of the potential to
experimental bulk quantities (see Sect.4 and Ref.\cite{Barreteau02}).
Note that in potentials
of the tight-binding type, the N-body part is strictly attractive
while the pairwise part is strictly repulsive.

Some physical insight can be gained by fixing the interatomic
distances to their bulk equilibrium values, i.e., ignoring atomic
relaxation effects. With this assumption $\sum_{j\neq i}
V(R_{ij})$ and $\sum_{j\neq i} g(R_{ij})$ are linear combinations
of the number of neighbors $Z^J_i$ of atom $i$ in the J$^{th}$
coordination sphere of radius $R_J$ ($R_J<R_c$) and
$E_i=E(Z^1_i...Z^J_i...)$. It is usual to take $R_1$ as the
reference distance and set $g(R_1)=1$. From the discussion of
Sect.5.1, we will first determine if there is any interaction
between steps. Obviously steps start to interact when the range of
potential is large enough. Then the two straight lines of Fig.\ref{fig:triangle}
transform into as many segments (with discontinuities of slopes)
as there are different step energies when $p$ increases.

\subsubsection{Pair potentials.}

These potentials are the simplest ones which have been used in the
past. We will limit ourselves to the study of unrelaxed surfaces
since it is well known that pair potentials most often lead to an outward
relaxation instead of the inward one generally observed at metal
surfaces. For any orientation of the surface it is easy to
determine the coordination numbers $Z^1_i, Z^2_i,...$for the
successive atomic layers $i$ from which the numbers $n^J_{step}$
are deduced. For the domain $(100)-(111)$ it is found that there
are no step interactions if $R_c<R_6$. Then $\Delta f(\eta)$ has
the triangular shape of Fig.\ref{fig:triangle} and its sign on the whole domain
$[0,\eta_c]$ is given by (see Eq.(\ref{eq:esc}))

\begin{equation}
\Delta E = E_S(311) - E_S(100) - E_S(111) = -4(V_3 + V_5)
\end{equation}

As a conclusion if $R_c<R_3$, $\Delta E=0$ so that the energy of
any vicinal surface is equal to the energy of the faceted
$(100)/(111)$ surface. If $R_c<R_6$, the surface is unstable if
$V_3+V_5<0$ and stable otherwise.

\subsubsection{N-body semi-empirical potentials.}

We now examine the case of semi-empirical potentials including an
N-body contribution and begin by neglecting atomic relaxation in
order to derive general trends for potentials of type
(\ref{eq:pot}). Then we will present a study of the stability of
Cu vicinal surfaces in the $(100)-(111)$ domain using the
semi-empirical potential set up in Sect.4.1.

When the interatomic distances are fixed to their bulk equilibrium
values, the energy of an atom $i$, $E_i(Z^1_i,...Z^J_i..)$, is no
longer a linear function of $Z^J_i$. However, as will be seen
below, the mathematical properties of the function has interesting
physical consequences. It can be shown easily that the step
energies of the $p(100)\times(111)$ and $p(111)\times(100)$ are
independent of $p$ as long as $R_c<R_3$ and are given by:

\begin{equation}
E^{p(100)\times(111)}_{step}= E(7,3) +E(10,5)-3E(8,5)/2-E(12,5)/2
\label{eq:escst100}
\end{equation}

\noindent and:

\begin{eqnarray}
\lefteqn{E^{p(111)\times(100)}_{step}=E(7,3)-5E(9,3)/3+E(10,5)+{}} \nonumber\\
& & {}+E(12,5)-4E(12,6)/3+{}
\label{eq:escst111}
\end{eqnarray}

Consequently for any semi-empirical potential of the form
(\ref{eq:pot}) including first and second nearest neighbors only,
$\Delta f(\eta)$ has a triangular shape when atomic relaxation is
neglected and its sign is given by:

\begin{equation}
\Delta E = E_S(311) - E_S(100) - E_S(111) \label{eq:deltae1a}
\end{equation}

or:

\begin{equation}
\Delta E = [E(7,3)+E(10,5)] - [E(8,5)+E(9,3)] \label{eq:deltae1b}
\end{equation}

Thus, in this approximation, $\Delta E$ arises from the difference
of the sum of energies of, on the one hand, atoms belonging to the
outer ($Z^1_i=7,Z^2_i=3$) and inner ($Z^1_i=10,Z^2_i=5$) step
edges, and on the other hand, of $(100)$($Z^1_i=8,Z^2_i=5$) and
$(111)$($Z^1_i=9,Z^2_i=3$) surface atoms.

As shown in Sect.5.2.1 the pair potential, when
limited to second nearest neighbors, does not contribute to
$\Delta E$. Noting that, since we have chosen $g(R_1)=1$,
$\rho_i=Z^1_i+Z^2_ig_2$ with $g_2=g(R_2)$, then we get:

\begin{equation}
\Delta E = [F(7+3g_2)-F(9+3g_2)]-[F(8+5g_2)-F(10+5g_2)]
\label{eq:deF1}
\end{equation}

For all the existing potentials of the form (\ref{eq:pot})
$F''(\rho)=d^2F/d\rho^2$ is positive. As a consequence
$F(\rho\!-\!2)\!-\!F(\rho)$ is a decreasing function of $\rho$,
therefore $\Delta E$ (and thus $\Delta f(\eta)$) is always
positive in the whole domain. This common property of this class
of potentials has a clear physical origin: the energy $E_i$ of an
atom $i$ should decrease more and more slowly when its
coordination increases towards the bulk coordination
\cite{Mcdsbook,Robertson91}. This clearly implies that $F''(\rho)$
must be positive. We have then proved that for {\it any} semi-empirical
potential of the general form (\ref{eq:pot}) on a rigid lattice at
0K and a cut-off radius $R_c<R_3$, {\it any}  metal vicinal surface between
(100) and (111) is {\it unstable} with respect to faceting.

So far we have demonstrated general results on the stability of
vicinal surfaces neglecting atomic relaxation. These results were
obtained under the assumption that the range of the potential is
restricted to the first two shells of neighbors. As shown in
Ref.\cite{Raouafi202} it is not possible to derive a general
behavior when the range of interactions is extended to further
neighbors. In order to investigate the effect of atomic relaxation
and of the range of interactions, it is necessary to have an
explicit expression of the potential. We will now present the
results obtained for Cu with the semi-empirical potential ($P_2$)
of Sect.4 whose range is limited to first and second neighbors. In
addition we have also considered the potential ($P_4$) of the same
type ($\alpha=2/3$) but with a cut-off radius between fourth and
fifth neighbors with parameters obtained from a least mean square
fit of the cohesive energy, the three elastic constants and the
bulk equilibrium distance. In all cases the atomic structure of
each vicinal surface has been fully relaxed using a conjugate
gradient algorithm.

\begin{figure}[!fht]
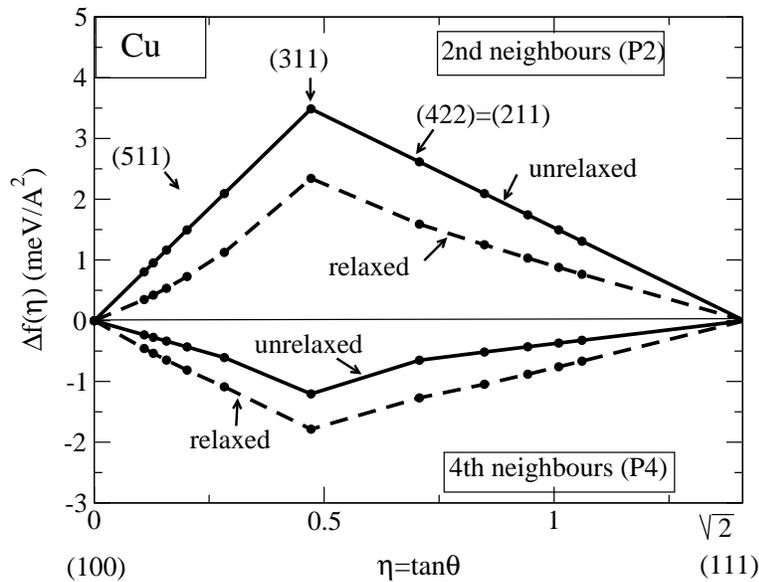

\begin{center}
  \ifpdf
  \includegraphics*[width=10cm]{figure11.pdf}
  \else
  \includegraphics*[width=10cm]{figure11.eps}
  \fi
\caption{$\Delta f(\eta)$ for Cu derived from the semi-empirical potentials
$P_2$ and $P_4$ given in the text corresponding to two cut-off radii with
and without relaxation for the (100) - (111) domain.}
\label{fig:deltaf_P2_P4}	  
\end{center}
\end{figure}

The results are given in Fig.\ref{fig:deltaf_P2_P4} for the $(100)-(111)$ domain
using both potentials and relaxed as well as unrelaxed surfaces. As
predicted from our previous analysis in the unrelaxed case $\Delta
f(\eta)$ has a triangular shape and is positive if the range of
the potential is restricted to second neighbors ($P_2$). However,
as expected, due to the effect of farther neighbors $\Delta
f(\eta)$ deviates from a triangular shape with potential $P_4$
and, more surprisingly, it changes sign. Thus all vicinal surfaces
between $(100)$ and $(111)$ become stable.

Finally, atomic relaxation always acts in favor of the
stabilization of vicinal surfaces since the atomic displacements
are larger on a vicinal surface than on a flat one (see Sect.4).
Nevertheless this effect is not large enough to modify the
stability or instability of the vicinal surfaces. Furthermore, as
clearly seen when using potential $P_2$ in the relaxed case,
$\Delta f(\eta)$ has no longer a triangular shape and is entirely
located above the triangle built from its tangent at both ends due
to the repulsive elastic step-step interactions (see Sect.4).

Let us discuss and summarize our results. From our analytical
study and Fig.\ref{fig:deltaf_P2_P4} it appears that the range of the
potential plays an important role but it is difficult to draw
general conclusions. In all cases considered here the effect of
farther neighbors is to act in favor of the stabilization of vicinal surfaces, however
including them will not automatically make vicinal surfaces
stable, this crucially depends on their relative importance and,
therefore, on the dependence  of the functions $V(r)$ and $g(r)$
with distance in (\ref{eq:pot}). The stability also depends on the
relative importance of $V$ with respect to $F(\rho)$ since, when
farther neighbors are included, both terms are present in the
energy balance. Moreover, in EAM and EMT potentials the N-body
and pair parts are not necessarily purely attractive or purely
repulsive, therefore even the sign of these terms is not known.
Let us finally compare our results with those of Frenken and
Stoltze \cite{Frenken99}. These authors have calculated $\Delta
f(\eta)$ for the fully relaxed $(100)$ and $(111)$ vicinal
surfaces of Ag (and other metals) using an EMT potential with
$R_3<R_c<R_4$ but in which the contribution of third neighbors is
nearly negligible. This explains the strong similarity between our
results on relaxed Cu (Fig. \ref{fig:deltaf_P2_P4}) with
potential $P_2$ and those of Frenken and Stoltze for Ag.

As a conclusion, the instability of vicinal surfaces at 0K claimed
by these authors is an {\it unavoidable} consequence of the type
of potential used when interactions are limited to first and
second neighbors.

\subsection{Stability of vicinal surfaces at 0K from tight-binding
calculations.}

It is interesting to deduce the function $\Delta f(\eta)$ from the
results presented in Sect.3. Indeed, the TB calculations are
certainly more realistic than those based on semi-empirical
potentials since they account for the influence of the angular
arrangement of neighbors and include electronic step-step
interactions (often oscillatory). We have seen in Sect.3 that
these interactions are small. However they may play a role in the
very delicate energy balance which governs the stability of
vicinal surfaces.

\begin{figure}[!fht]
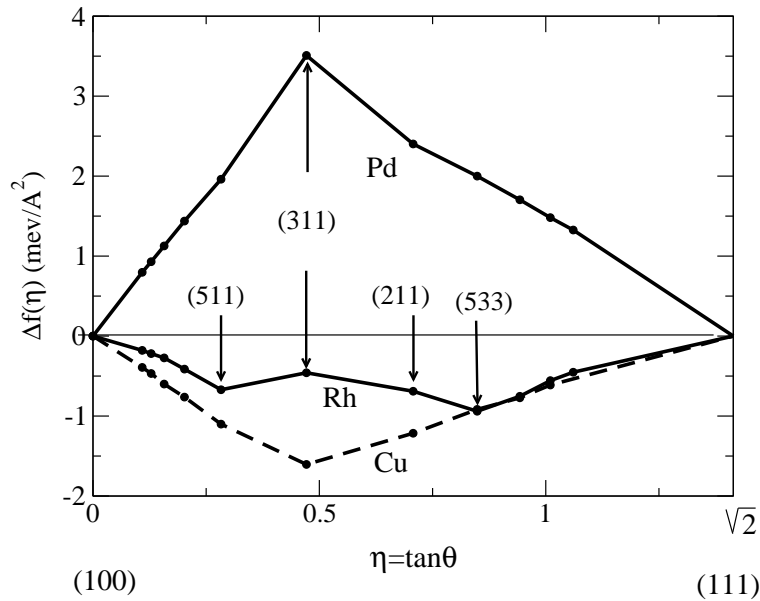

\begin{center}
  \ifpdf
  \includegraphics*[width=10cm]{figure12.pdf}
  \else
  \includegraphics*[width=10cm]{figure12.eps}
  \fi
\caption{$\Delta f(\eta)$ for Rh, Pd and Cu from tight-binding calculations
for the (100) - (111) domain.}
\label{fig:RhPdCuTB}	  
\end{center}
\end{figure}

The functions $\Delta f(\eta)$ for Rh, Pd and Cu in the
$(100)-(111)$ domain are plotted in Fig.\ref{fig:RhPdCuTB} and
show very different behaviors depending on the chemical element.
For Cu all vicinal surfaces in the domain are stable at 0K while
for Pd they are unstable. Rh behaves quite differently: even
though all vicinal surfaces are stable with respect to faceting
into $(100)$ and $(111)$ facets, the vicinal surfaces of
orientation such that $\sqrt(2)/5<\eta<3\sqrt(2)/5$ are unstable
relative to faceting into $(511)$ and $(533)$ orientations which
correspond to the two local minima in $\Delta f(\eta)$. This
peculiar behavior is clearly related to the electronic step-step
interactions which are repulsive for the $(311)$ and $(211)$
surfaces and attractive for $(511)$ and $(533)$ surfaces (see
Fig.\ref{fig:estepp}).

\subsection{Finite temperature effects.}

So far all calculations were carried out at 0K. Therefore it is
important to know whether the effect of a finite temperature may
be large enough to reverse the stability of vicinal surfaces with
respect to faceting. The variation of $f(\eta)$ with temperature
arises from two contributions: the vibrational effects that have
been studied in Sect.4 and the meandering of steps, which is
regulated by the kink formation energy, giving rise to entropy
contributions. Let us first discuss the order of magnitude of the
latter. In the limit of infinite terraces steps fluctuate
independently of each other but when the terrace width decreases
the entropy gain due to the meandering is limited by the
non-crossing condition which gives rise to a repulsive step-step
interaction. Actually, both effects are driven by a parameter
$\zeta=exp(-\epsilon_{kink}/k_BT)$ where $\epsilon_{kink}$ is the
kink formation energy. As can be seen from Table 4 this parameter
is quite small, at least up to room temperature, and, as a
consequence,these two contributions are negligible compared with
the value of $\Delta f(\eta)$ at 0K and 
they are also small compared with the contribution $\Delta
f_{vib}(\eta)$ due to vibrations \cite{Raouafi202}.

\begin{figure}[!fht]
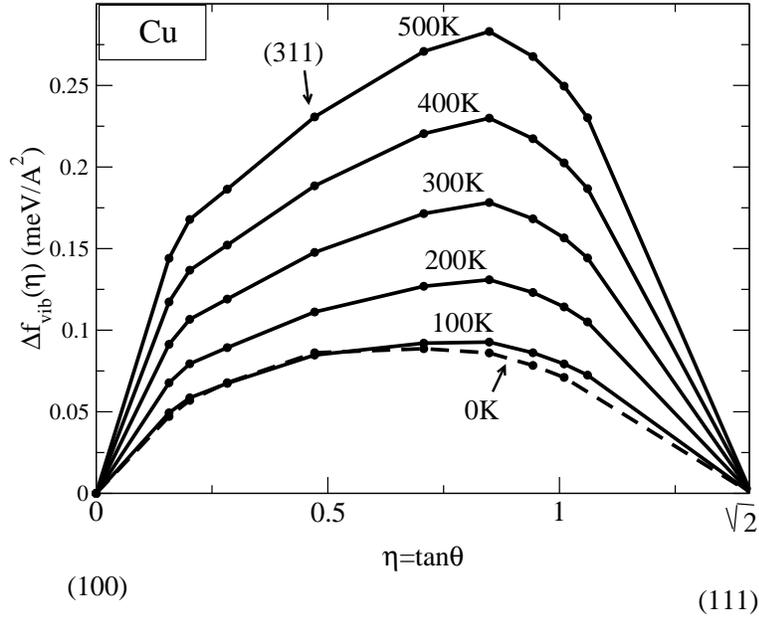

\begin{center}
  \ifpdf
  \includegraphics*[width=10cm]{figure13.pdf}
  \else
  \includegraphics*[width=10cm]{figure13.eps}
  \fi
\caption{$\Delta f_{vib}(\eta)$ in the (100) - (111) domain from potential $P_2$
as a function of temperature.}
\label{fig:deltafvib}	  
\end{center}
\end{figure}

From the study of vibrations presented in Sect.4, we can easily
derive $\Delta f_{vib}(\eta)$ as a function of temperature 
in the $(100)-(111)$ domain. The
corresponding curves for copper is drawn in Fig.\ref{fig:deltafvib}. It can be seen
that $\Delta f_{vib}$ is positive but its order of magnitude is
not large enough to destabilize the vicinal surfaces of Cu in this
domain. These results are in contradiction with those of Frenken
and Stoltze \cite{Frenken99}. Actually these authors evaluated the
vibrational entropy contribution using a simplified Einstein model
and neglected the vibrational internal energy which is justified
at room temperature, but not at low temperature. Moreover they only
included the perturbation between the outer edge and a $(111)$
surface atom and not the term coming from the perturbation between
the inner edge and the $(100)$ surface atom (see Eq.(\ref{eq:deltae1b})). These two terms are
of opposite sign and are expected to be of the same order of
magnitude. Consequently not only the estimate of $\Delta f_{vib}$
in \cite{Frenken99} is too large but even the sign is wrong.

In conclusion, the contribution of vibrations $\Delta f_{vib}$ to $\Delta f$
is quite small. Thus its calculation needs a precise knowledge of
the phonon spectra including both vibrational internal energy and
entropy, at least if the temperature varies from 0K to $\simeq$
300K.

\section{Conclusion}

In conclusion, with the {\it spd} tight-binding method it is
possible to carry out realistic calculations on the energetics of
vicinal surfaces: surface, step and kink energies. Indeed, the
formation energies of isolated steps, obtained with this method, on
Rh, Pd and Cu surfaces for various geometries are in good
agreement with existing experimental data. In particular, our
results predict that an adisland of Cu on Cu(100) should be a
square with broken corners at 0K. Moreover the correct relative
stability of the two types of steps (A and B) on the vicinal
surfaces of Cu(111) is obtained. Kink energies have also been
calculated and compare nicely with experimental data.

Following the approach of Vitos et al \cite{Vitos99} and from the
knowledge of the surface energies of the three low index surfaces
calculated from the $spd$ TB hamiltonian, effective pair
interactions can be deduced giving step and kink energies in good
agreement with those derived from the diagonalization of the same
hamiltonian for vicinal surfaces. However in this approach
step-step interactions are disregarded. On the contrary, our
study, based on the calculation of step energies on vicinal
surfaces as a function of the terrace width, has enabled us to
derive electronic step-step interactions for narrow and moderately wide
terraces (d $\leq$ 20{\AA}). These interactions are rapidly
decaying and they may be attractive or repulsive depending on the
terrace width. Moreover, in this range of widths, their order of
magnitude is comparable to that of other interactions.

A semi-empirical potential for Copper, including an N-body
contribution, has been built. It accounts for the multilayer
relaxation of vicinal surfaces and describes accurately their
localized vibration modes observed in IHAS and EELS. The
contribution of vibrations to the free energy of steps has also
been calculated as a function of the distance between steps and it
was found that the step-step interactions mediated by phonons are
attractive for Cu vicinal surfaces in the (100)-(111) domain.

The stability of vicinal surfaces with respect to faceting has
also been investigated. The conclusions derived from semi-empirical
potentials have been criticized. Contrary to the results obtained
from these potentials which predict that vicinal surfaces of
metals are unstable at 0K, the tight-binding electronic structure
calculations lead to a variety of behaviors: a vicinal surface in
the (100)-(111) domain may be stable (Cu) or unstable (Pd)
relative to faceting into (100) and (111) facets or may even
undergo a faceting towards other vicinal surfaces (Rh). Finally,
temperature effects have been found to be negligible for Cu, at
least up to room temperature.

\newpage

\newpage

\end{document}